%% file: jakiupdatev1.tex
\documentclass[aps,prc,preprintnumbers,superscriptaddress,floatfix,showpacs,preprint]{revtex4-1}
\usepackage{epsfig}
\usepackage{dcolumn}
\usepackage{bm}
\usepackage{amssymb}
\usepackage{amsmath}

\begin{document}
\title{Bulk viscosity-driven suppression of shear viscosity effects on the flow harmonics at RHIC}

\author{Jacquelyn Noronha-Hostler}
\affiliation{Instituto de F\'{i}sica, Universidade de S\~{a}o Paulo, C.P.
66318, 05315-970 S\~{a}o Paulo, SP, Brazil}

\author{Jorge Noronha}
\affiliation{Instituto de F\'{i}sica, Universidade de S\~{a}o Paulo, C.P.
66318, 05315-970 S\~{a}o Paulo, SP, Brazil}

\author{Fr\'ed\'erique Grassi}
\affiliation{Instituto de F\'{i}sica, Universidade de S\~{a}o Paulo, C.P.
66318, 05315-970 S\~{a}o Paulo, SP, Brazil}

\date{\today}
\begin{abstract}
The interplay between shear and bulk viscosities on the flow harmonics, $v_n$'s, at RHIC is investigated using the newly developed relativistic 2+1 hydrodynamical code v-USPhydro that includes bulk and shear viscosity effects both in the hydrodynamic evolution and also at freeze-out.  While shear viscosity is known to attenuate the flow harmonics, we find that the inclusion of bulk viscosity decreases the shear viscosity-induced suppression of the flow harmonics bringing them closer to their values in ideal hydrodynamical calculations. Depending on the value of the bulk viscosity to entropy density ratio, $\zeta/s$, in the quark-gluon plasma, the bulk viscosity-driven suppression of shear viscosity effects on the flow harmonics may require a re-evaluation of the previous estimates of the shear viscosity to entropy density ratio, $\eta/s$, of the quark-gluon plasma previously extracted by comparing hydrodynamic calculations to heavy ion data. 
\end{abstract}
\maketitle

\section{Introduction}

One of the the main results stemming from heavy-ion collision experiments at the Relativistic Heavy Ion Collider (RHIC) and the Large Hadron Collider (LHC) is the discovery that the Quark-Gluon Plasma (QGP) behaves as a nearly perfect fluid \cite{Gyulassy:2004zy} in which the lowest value of $\eta/s \sim 0.2$ \cite{Heinz:2013th}.  While there have been several model calculations that support such a small value for $\eta/s$ in the QGP \cite{Danielewicz:1984ww,Kovtun:2004de,Nakamura:2004sy,Meyer:2007ic,Xu:2007ns,Xu:2007jv,Hirano:2005wx,Csernai:2006zz,Hidaka:2008dr,NoronhaHostler:2008ju,NoronhaHostler:2012ug,Ozvenchuk:2012kh}, much less is known about the bulk viscosity to entropy density ratio, $\zeta/s$. In fact, though it is true that $\zeta/s$ vanishes at sufficiently large temperatures \cite{Arnold:2006fz}, it is not clear at the moment how large this quantity can be  \cite{Meyer:2007dy,Karsch:2007jc} in the range of temperatures probed in heavy ion collisions, $T \sim 100-400$ MeV.

This has led to the idea that the bulk viscosity may be extremely small and have negligible effects on observable quantities such as the azimuthal flow anisotropies (for phenomenological consequences of large bulk viscosity in heavy ion collisions see \cite{Torrieri:2007fb,Rajagopal:2009yw,Habich:2014tpa}).  Most studies have used only shear viscous calculations and then fitted the calculated flow harmonics to experimental data to estimate the shear viscosity of the QGP \cite{etas}. There are a few exceptions of those who have explored bulk viscosity \cite{Song:2009rh,Bozek:2009dw,Monnai:2009ad,Denicol:2009am,Denicol:2010tr,Dusling:2011fd} and its effects on elliptic flow but further work was needed to quantify the effects of bulk viscosity on the higher order flow harmonics. Recently, in \cite{Noronha-Hostler:2013gga} the effects solely from bulk viscosity on the flow harmonics at RHIC were investigated using event-by-event hydrodynamics and it was found that bulk viscosity enhances the differential flow harmonics with respect to the ideal case, which is the opposite effect found in the case of shear viscosity \cite{etas}.

In this paper we will explore the interplay between bulk and shear viscosities within the framework of relativistic hydrodynamical modeling using v-USPhydro \cite{Noronha-Hostler:2013gga}, which is a boost invariant viscous hydrodynamical code that runs event-by-event initial conditions using Smoothed Particle Hydrodynamics (SPH) \cite{originalSPH,SPHothers,Aguiar:2000hw} to solve the equations of motion. For some choices of the model parameters, we find that for $\sqrt{s}=200$ A GeV RHIC collisions bulk viscosity can almost entirely negate the effects of shear viscosity when they are of a comparable size for both the integrated and $p_T$ dependent flow harmonics. In fact, in this case for differential flow harmonics bulk viscosity effects dominate over the effects from shear.  However, we find that there is a strong dependence on the model choice of bulk viscous corrections at freeze-out for the differential flow harmonics at intermediate $p_T> 1.5$ GeV (at low $p_T$ both methods converge and, thus, the integrated flow harmonics are much more robust with respect to model changes in the viscous contribution to the particle distributions).  Finally, we find that bulk viscosity has a nontrivial effect on the shear stress tensor even when the chosen $\zeta/s$ is significantly smaller than the shear viscosity.    

This paper is organized as follows. In Section \ref{sec:model} we cover the relativistic hydrodynamical model that we are using.  In \ref{sec:eom} we discuss the equations of motion for 2+1 relativistic hydrodynamics with bulk and shear viscosity using the SPH formalism then in Section \ref{sec:bulkshear} we show the transport coefficients used for this paper.  In Section \ref{sec:ic} the setup for our Glauber event-by-event initial conditions are discussed and in Section \ref{sec:freezeout} we discuss the parameters and equations for the freeze-out with viscous corrections.  In Section \ref{sec:visceffhydro} we explore the effects of shear and bulk viscosities on the hydrodynamical evolution while in Section \ref{sec:results} we discuss the results of our work for both the integrated $v_n$'s in \ref{sec:intvns} and the differential $v_n$'s in \ref{eqn:difvns}.  We also show a comparison for the case when bulk and shear viscosities have the same magnitude in Section \ref{eqn:equal}.  Finally, in Section \ref{sec:conclu} we discuss the consequences of our work. Details about the equations and tests of the accuracy of v-USPhydro can be found in the Appendices.

\textit{Definitions}:  We use a flat space-time metric in Milne coordinates defined as $g_{\mu \nu }=(1,-1,-1,-\tau ^{2})$ where $x^{\mu
}=(\tau ,\mathbf{r},\eta )$ and 
\begin{eqnarray}
\tau &=&\sqrt{t^{2}-z^{2}}  \nonumber \\
\eta &=&\frac{1}{2}\ln \left( \frac{t+z}{t-z}\right) \,.
\end{eqnarray}  
are the proper time and space-time rapidity, respectively. Furthermore, we assume boost invariance for the flow velocity so $u_{\mu }=\left( 
\sqrt{1+u_{x}^{2}+u_{y}^{2}},u_{x},u_{y},0\right) $ and also $u_\mu u^\mu=1$. Natural units are employed throughout this work, i.e., $\hbar=k_B=c=1$. 

\section{Details of the Hydrodynamic Model}\label{sec:model}

\subsection{Equations of Motion for 2+1 relativistic hydrodynamics with bulk and shear viscosities}\label{sec:eom}
In this paper we use a boost invariant setup with a vanishing baryon chemical potential and the conservation of energy and momentum $\nabla_\mu T^{\mu\nu}=0$ can be written as
\begin{equation}
\frac{1}{\tau}\partial _{\mu }\left( \tau T^{\mu \nu }\right)
+\Gamma _{\lambda \mu }^{\nu }T^{\lambda \mu }=0  \label{eqn:hydro}
\end{equation}%
where the Christoffel symbol is 
\begin{equation}
\Gamma _{\lambda \mu }^{\nu }=\frac{1}{2}g^{\nu \sigma }\left( \partial
_{\mu }g_{\sigma \lambda }+\partial _{\lambda }g_{\sigma \mu }-\partial
_{\sigma }g_{\mu \lambda }\right) .
\end{equation}%
The general expression for energy-momentum tensor that includes both bulk and shear viscosity effects is
\begin{equation}
T^{\mu \nu }=\varepsilon u^{\nu}u^{\nu }-\left( p+\Pi \right) \Delta ^{\mu
\nu }+\pi^{\mu\nu},
\end{equation}%
where $\Pi$ is the bulk viscous pressure, $\pi^{\mu\nu}$ is the shear stress tensor, and the spatial projection operator is $%
\Delta _{\mu \nu }=g_{\mu \nu }-u_{\mu }u_{\nu }$.  Above, we use the Landau
definition for the local rest frame, $u_{\nu }T^{\mu \nu }=\varepsilon
u^{\mu }$ and the remaining dynamical quantities are the energy density $%
\varepsilon$, the pressure $p$ (which can be written in terms of $\varepsilon$ via the equation of state) and the fluid 4-velocity $u^{\mu}$. The dissipative currents $\Pi$ and $\pi^{\mu\nu}$ obey relaxation-type differential equations and the full form of these equations, at least according to kinetic theory, can be found in \cite{Denicol:2012cn}. In this paper we do not consider all the terms found in \cite{Denicol:2012cn} due to the large uncertainty regarding the values of the many new transport coefficients involved (for a recent study involving the determination of these coefficients in certain limits see \cite{Denicol:2014vaa}). Rather, we use as in our previous work \cite{Noronha-Hostler:2013gga} the simplest equation for the bulk scalar (obtained originally via the memory function prescription \cite{Koide:2006ef} and used also in \cite{Denicol:2009am,Denicol:2010tr}) 
\begin{equation}
\tau _{\Pi }\left( D\Pi +\Pi \theta \right) +\Pi +\zeta \theta =0,
\label{eqn:hydro3}
\end{equation}%
where $D=u^{\mu }\nabla_{\mu }$ is the comoving covariant derivative, $%
\theta \equiv \nabla_\mu u^\mu=\tau ^{-1}\partial _{\mu }\left( \tau u^{\mu }\right) $ is the fluid
expansion rate, and $\tau _{\Pi}$ is the
bulk relaxation time coefficient. For the description of the shear stress tensor we use the minimal Israel-Stewart description (compatible with conformal invariance)
\begin{equation} 
\tau_{\pi}\left(\Delta_{\mu\nu\alpha\beta}D\pi^{\alpha\beta}+\frac{4}{3}\pi_{\mu\nu}\theta\right)+\pi_{\mu\nu}=2\eta\sigma_{\mu\nu}
\end{equation}
where we defined the tensor projector $\Delta_{\mu\nu\alpha\beta}=\frac{1}{2}\left[\Delta_{\mu\alpha}\Delta_{\nu\beta}+\Delta_{\mu\beta}\Delta_{\nu\alpha}-\frac{2}{3}\Delta_{\mu\nu}\Delta_{\alpha\beta}\right]$, the shear tensor $\sigma_{\mu\nu}=\Delta_{\mu\nu\alpha\beta}\nabla^\alpha u^\beta$, and $\tau_{\pi}$ is the shear relaxation coefficient. Therefore, in this work we have 4 transport coefficients: $\zeta$, $\eta$ and their respective relaxation times $\tau_{\Pi}$ and $\tau_{\pi}$. We note that we included the term $\pi^{\mu\nu}\theta$ in the equations of motion for $\pi^{\mu\nu}$ to make it possible to check the accuracy of our code against the analytical and semi-analytical solutions found in Ref.\ \cite{Marrochio:2013wla} (shown in detail in Appendix \ref{sheartest}).

The fluid dynamical evolution is written using the Lagrangian approach within the Smoothed Particle Hydrodynamics (SPH).  An in depth discussion of the SPH formalism and its relationship to the equations of motion can be found in \cite{Hama:2004rr,Denicol:2009am,Denicol:2010tr,Noronha-Hostler:2013gga,Noronha-Hostler:2013ria,Andrade:2013poa}.

\subsection{Model choice for the transport coefficients and equation of state}\label{sec:bulkshear}

\begin{figure}[ht]
\includegraphics[width=0.4\textwidth]{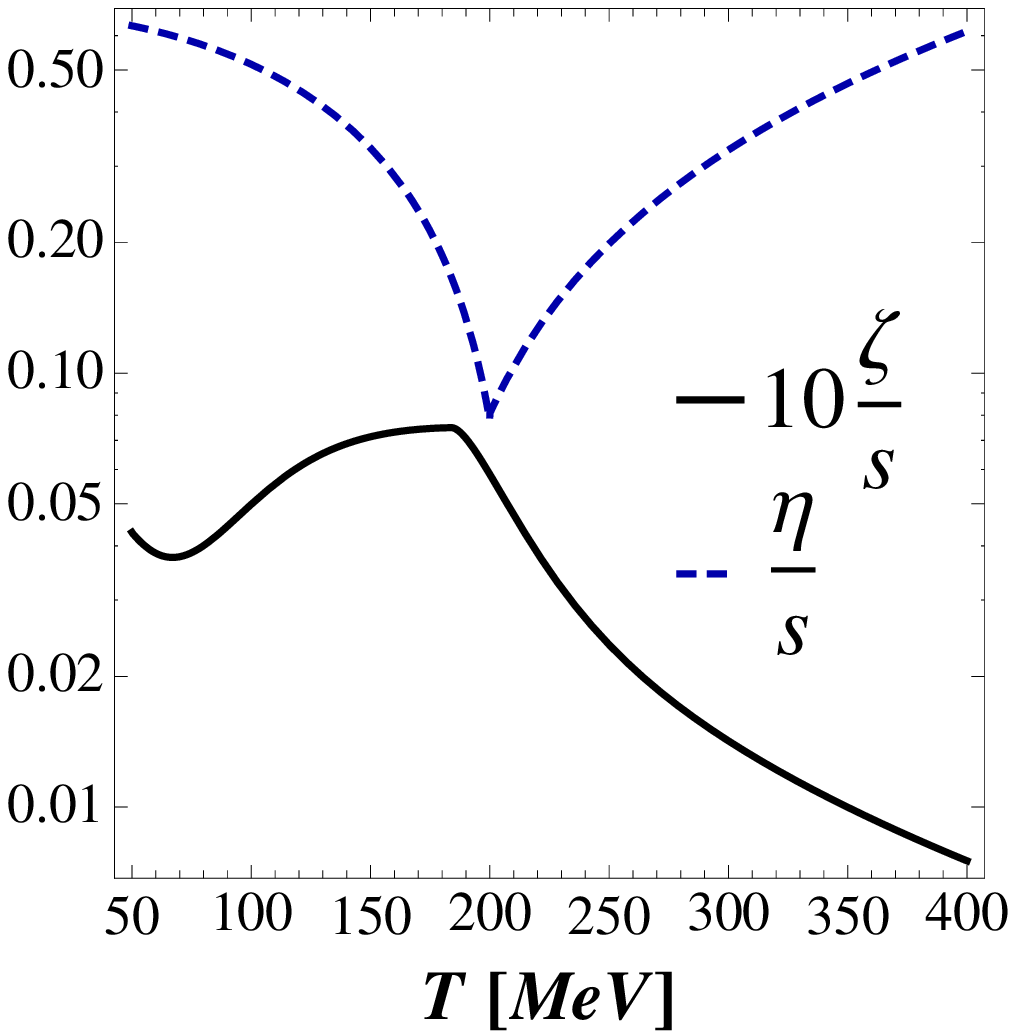} \\
\includegraphics[width=0.4\textwidth]{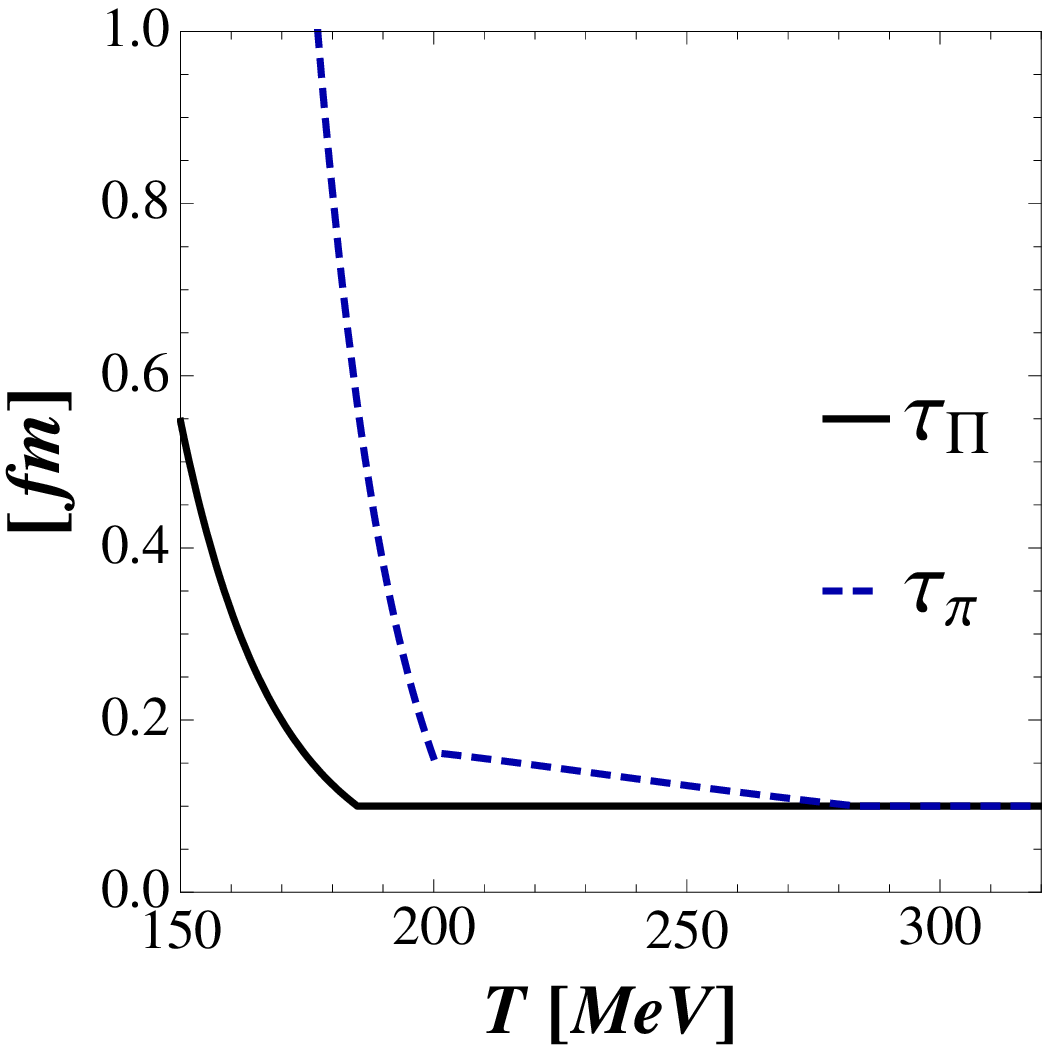} \\
\caption{(Color online) Upper panel - Temperature dependence of $\eta/s$ from Eq.\ (\ref{eqn:eta}) (dashed blue line) and $\zeta/s$ (multiplied by a factor of 10 for clarity) obtained from Eq.\ (\ref{eqn:adszeta}) (solid black line). Lower panel - The relaxation time coefficients $\tau_{\pi}$ from Eq.\ (\ref{eqn:taupi}) for shear (dashed blue line) and $\tau_{\Pi}$ for bulk from Eq.\ (\ref{eqn:tauPI}) (solid black line).}
\label{fig:transco}
\end{figure}

The v-USPhydro code has the ability to run ideal, bulk, shear, and shear+bulk 2+1 hydrodynamics (a generalization of the code to include full 3+1 dynamics is in the making).  In this paper we consider the temperature dependent shear, bulk, and relaxation time coefficients shown in Fig.\ \ref{fig:transco}.

For the temperature dependent shear viscosity we use the parametrization done in \cite{Niemi:2012ry} that describes the low temperature region using the result from the extended mass spectrum hadronic model \cite{NoronhaHostler:2008ju} while at high temperatures $\eta/s$ is given by the lattice data of Ref.\ \cite{Nakamura:2004sy}. It reads
\begin{eqnarray}
\frac{\eta}{s}(T>T_{tr})&=&-0.289+0.288\left(\frac{T}{T_{tr}}\right)+0.0818\left(\frac{T}{T_{tr}}\right)^2\nonumber\\
\frac{\eta}{s}(T<T_{tr})&=&0.681-0.0594\left(\frac{T}{T_{tr}}\right)-0.544\left(\frac{T}{T_{tr}}\right)^2\label{eqn:eta}
\end{eqnarray}
where $T_{tr}=180$ MeV and the shear relaxation time \cite{Denicol:2010xn,Denicol:2011fa} is taken to be
\begin{equation}\label{eqn:taupi}
\tau_{\pi}=5\eta/(\varepsilon+p).
\end{equation}

We have used the following bulk viscosity coefficient (inspired by Buchel's formula \cite{Buchel:2007mf} for a strongly coupled plasma) 
\begin{equation}
\frac{\zeta }{s}=\frac{1}{8\pi }\left( \frac{1}{3}-c_{s}^{2}\right) ,
\label{eqn:adszeta}
\end{equation}%
with the corresponding temperature dependent bulk relaxation time, $\tau _{\Pi }$ (see \cite{Huang:2010sa})
\begin{equation}
\tau _{\Pi }=9\,\frac{\zeta }{\varepsilon -3p}\,.  \label{eqn:tauPI}
\end{equation}%
Given the small value of $\zeta/s$ used here, we note that in Fig.\ \ref{fig:transco} we actually plot $10\,\zeta/s$ in order to better illustrate its temperature dependence. Furthermore, we always ensure that $\tau _{\Pi }$ and $\tau_{\pi}$ are greater than 0.1 fm (the time step size of the numerical code) to avoid stability issues. 

\subsection{Initial conditions}\label{sec:ic}

\begin{table}
\begin{center}
 \begin{tabular}{|c|c|}
 \hline
 Centrality & $N_{part}$  \\
 \hline
 $0-10\%$ & $>$ 274.95     \\
 $10-20\%$ & 195.98-274.95     \\
 $20-30\%$ & 139.01-195.98     \\
 $30-40\%$ & 96.99-139.01     \\
 $40-50\%$ & 61.95-96.99     \\
 $50-60\%$ & 37.04-61.95     \\
 \hline
 \end{tabular}
 \end{center}
 \caption{Relationship between the number of participants, $N_{part}$, and the different centrality classes for Au+Au collisions at RHIC at $\sqrt{s}_{NN}=200$ GeV used in this paper.}
 \label{tab:par}
 \end{table}

In this paper we only consider  Monte Carlo Glauber simulations of Au+Au collisions at RHIC at $\sqrt{s}_{NN}=200$ GeV \cite{ic}  as our initial conditions for the energy density.  We begin the relativistic fluid-dynamical evolution at $\tau _{0}=1$ fm (for testing of this assumption see \cite{Noronha-Hostler:2013gga}).  Our centrality classes are found by binning the results for $N_{part}$ over 15,000 events and they are well in agreement with other Monte Carlo Glauber simulations \cite{Adler:2003cb}. The relationship between $N_{part}$ and the centrality classes is shown in Tab.\ \ref{tab:par}.  Within each centrality class we calculate 150 hydrodynamical events on an event-by-event basis.  

As in our previous work using the v-USPhydro code \cite{Noronha-Hostler:2013gga}, our initial energy density is
\begin{equation}\label{eqn:cglauber}
\varepsilon (\mathbf{r})=c\;n_{coll}(\mathbf{r}),
\end{equation}%
where $n_{coll}$ is the number density of binary collisions in the event, which is fixed to obtain on average $123$ direct $\pi ^{+}$'s in central (averaged $0-5\%$) RHIC collisions (this number of direct pions, when added to the yield coming from particle decays, leads to the correct number of $\pi ^{+}$'s in this case). Also, in this paper particle decays and hadronic transport have not been taken into account. Furthermore, we assume that $\Pi$, $\pi^{\mu\nu}$, and the spatial components of $u^{\mu}$ vanish at $\tau_{0}$.

\subsection{Cooper-Frye Freeze-out}\label{sec:freezeout}

Viscous corrections enter not only in the hydrodynamical equations of motion discussed in Section \ref{sec:eom} but also in the distribution function for the Cooper-Frye freeze-out \cite{Cooper:1974mv}. We perform the freeze-out on an isothermal hypersurface with the freeze-out temperature $T_0=150$ MeV \cite{Noronha-Hostler:2013gga}. The distribution function for a given hadron is described as
\begin{equation}
f_{p}=f_{0p}\left\{1+\left(1-af_{0p}\right)\left[\delta f^{Bulk}_{p}+\delta f^{Shear}_{p}\right]\right\}
\end{equation}
where the ideal component of the distribution function, $ f_{0p}$, is
\begin{equation}\label{eqn:CFideal}
f_{0p}=\frac{1}{e^{(p^{\mu}u_{\mu})/T_0}+a}
\end{equation}
where $a=1$ for fermions, $a=-1$ for bosons, and $a=0$ for classical Boltzmann statistics.
The general form of the correction term for bulk viscosity, $\delta f^{Bulk}_{p}$, up to second order in powers of $\left(u^i\cdot p_i\right)$ is \cite{Noronha-Hostler:2013gga}
\begin{equation}
\delta f^{Bulk}_{p}=\Pi\left[B_0+D_0\left(u^i\cdot p_i\right)+E_0\left(u^i\cdot p_i\right)^2\right]
\end{equation}
where $B_0$, $D_0$, and $E_0$ depend on the particle type (mass, degeneracy) and freeze-out temperature.  In this paper we consider both the coefficients determined from the Moments Method (MOM) in \cite%
{Denicol:2012cn,Denicol:2012yr,Noronha-Hostler:2013gga} and those derived by Monnai and Hirano (MH) in \cite{Monnai:2009ad}. MH implemented Grad's 14-moment method for multi-particle species to compute the bulk viscous contribution to the distribution function. MOM is based on the novel procedure proposed in \cite{Denicol:2012cn} to derive viscous hydrodynamic equations from the Boltzmann equation, generalized to include the case involving different hadron species.

The exact coefficients for each method were determined in \cite{Noronha-Hostler:2013gga} for the case of pions with $T_0=150$ MeV and for MOM we obtain
\begin{eqnarray}
B_{0}^{(\pi)}&=& -65.85 \,\,fm^4\,,  \nonumber \\
D_{0}^{(\pi)}&=& 171.27 \,\,fm^4/GeV\,,  \nonumber \\
E_{0}^{(\pi)}&=& -63.05 \,\,fm^4/GeV^2\,,
\end{eqnarray}
while for MH
\begin{eqnarray}
B_{0}^{(\pi)}&=& -0.69 \,\,fm^4\,,  \nonumber \\
D_{0}^{(\pi)}&=& -38.96 \,\,fm^4/GeV\,,  \nonumber \\
E_{0}^{(\pi)}&=& 49.69 \,\,fm^4/GeV^2\,.
\end{eqnarray}
 
Finally, we take the ``democratic" Ansatz for the correction term for shear viscosity, $\delta f^{Shear}_{p}$
\begin{eqnarray}
\delta f^{Shear}_{p}&=&\frac{\pi^{\mu\nu}p_{\mu}p_{\nu}}{2\left(\varepsilon+P\right)T^2}
\end{eqnarray}
(for a recent discussion on the validity of such an Ansatz in kinetic theory see \cite{Molnar:2014fva}). 

The final expression for the pion spectrum in the SPH formalism, including both shear and bulk viscosity effects, is worked out in Appendix \ref{detailsCF} and we refer the reader to that section for the necessary details.

\section{Viscous Effects in the Hydrodynamical Evolution}\label{sec:visceffhydro}

\begin{figure}[ht]
\centering
\begin{tabular}{cc}
\includegraphics[width=0.5\textwidth]{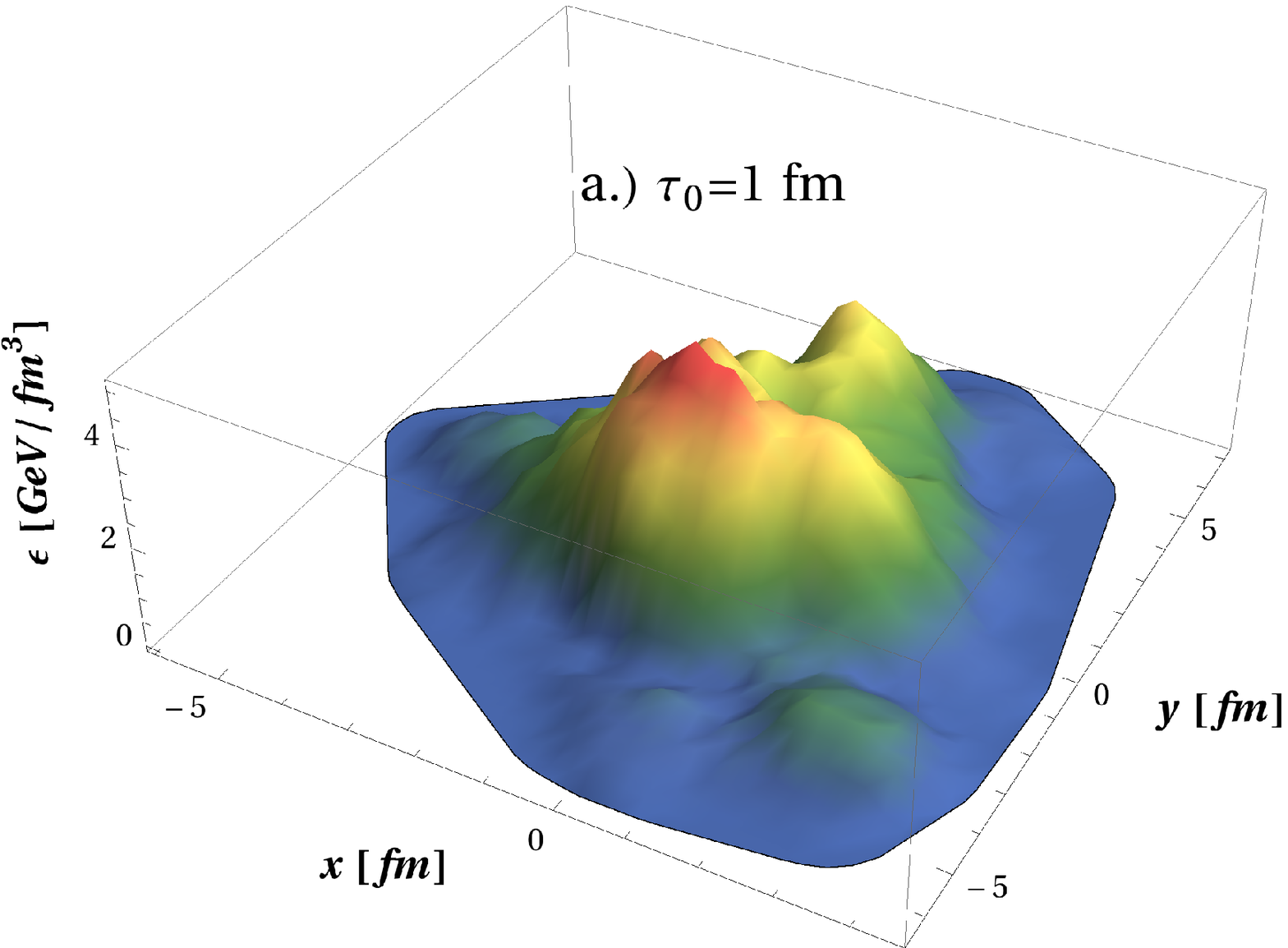}  &  \\
\includegraphics[width=0.5\textwidth]{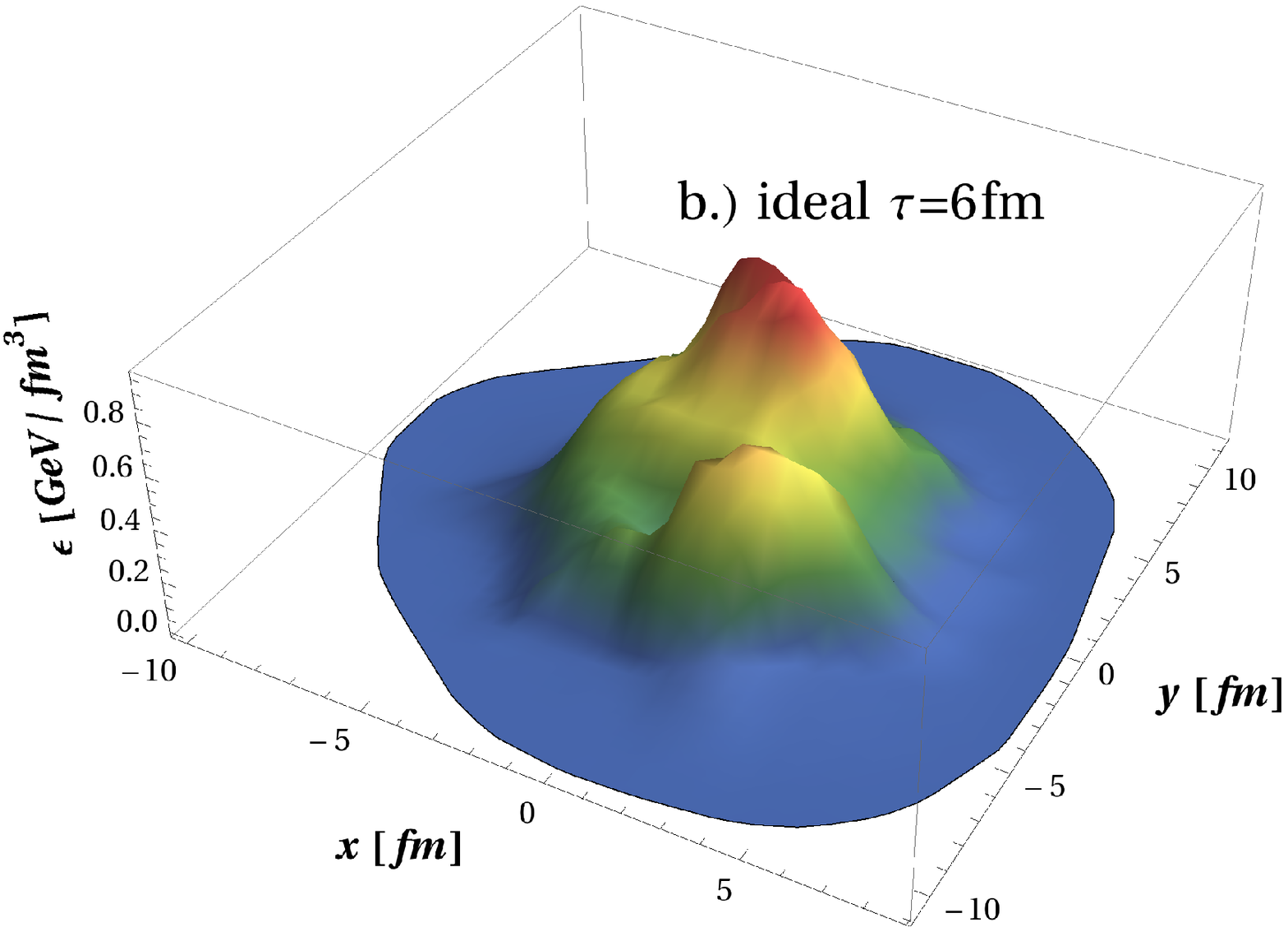} & 
\includegraphics[width=0.5\textwidth]{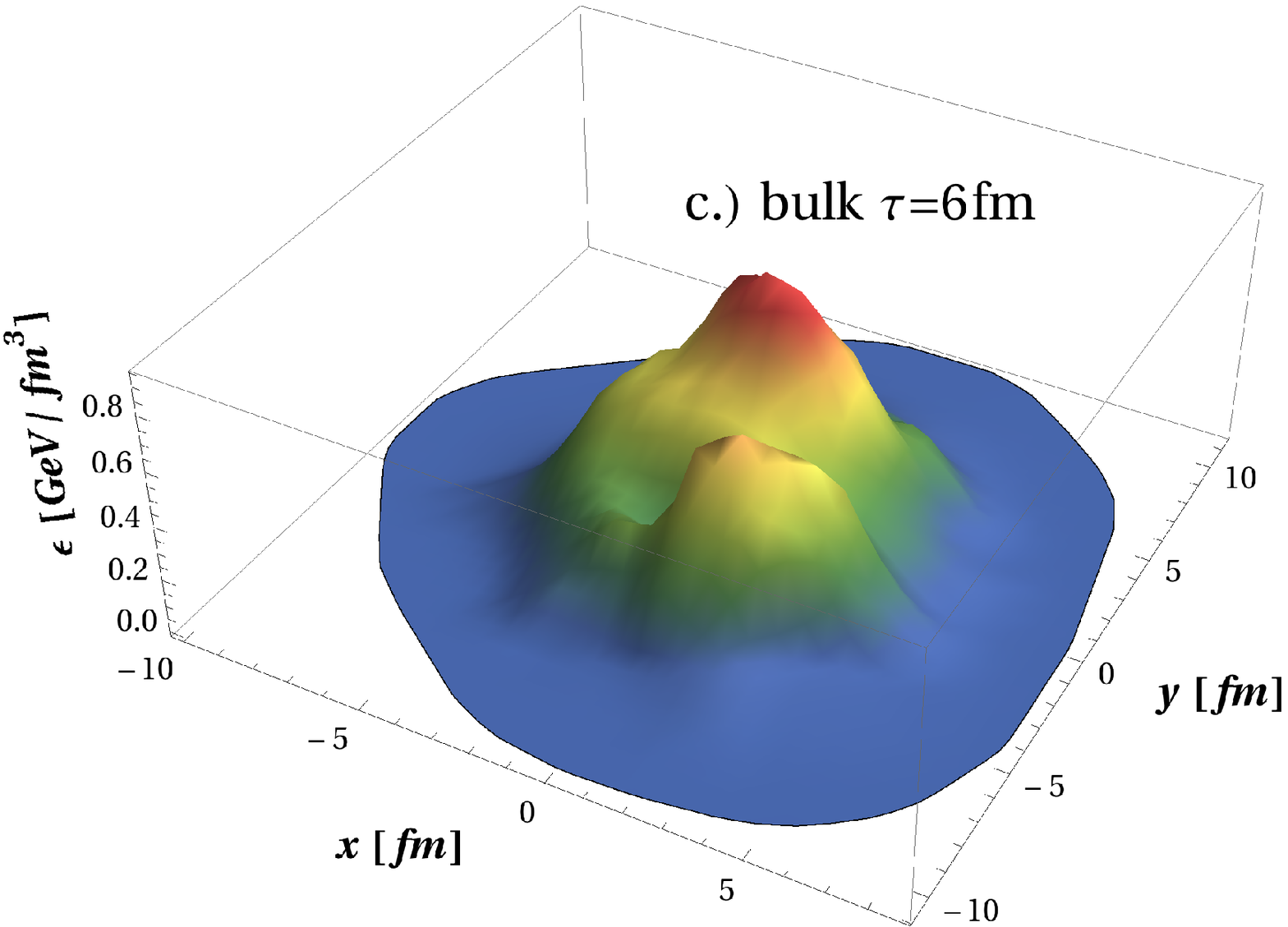} \\
\includegraphics[width=0.5\textwidth]{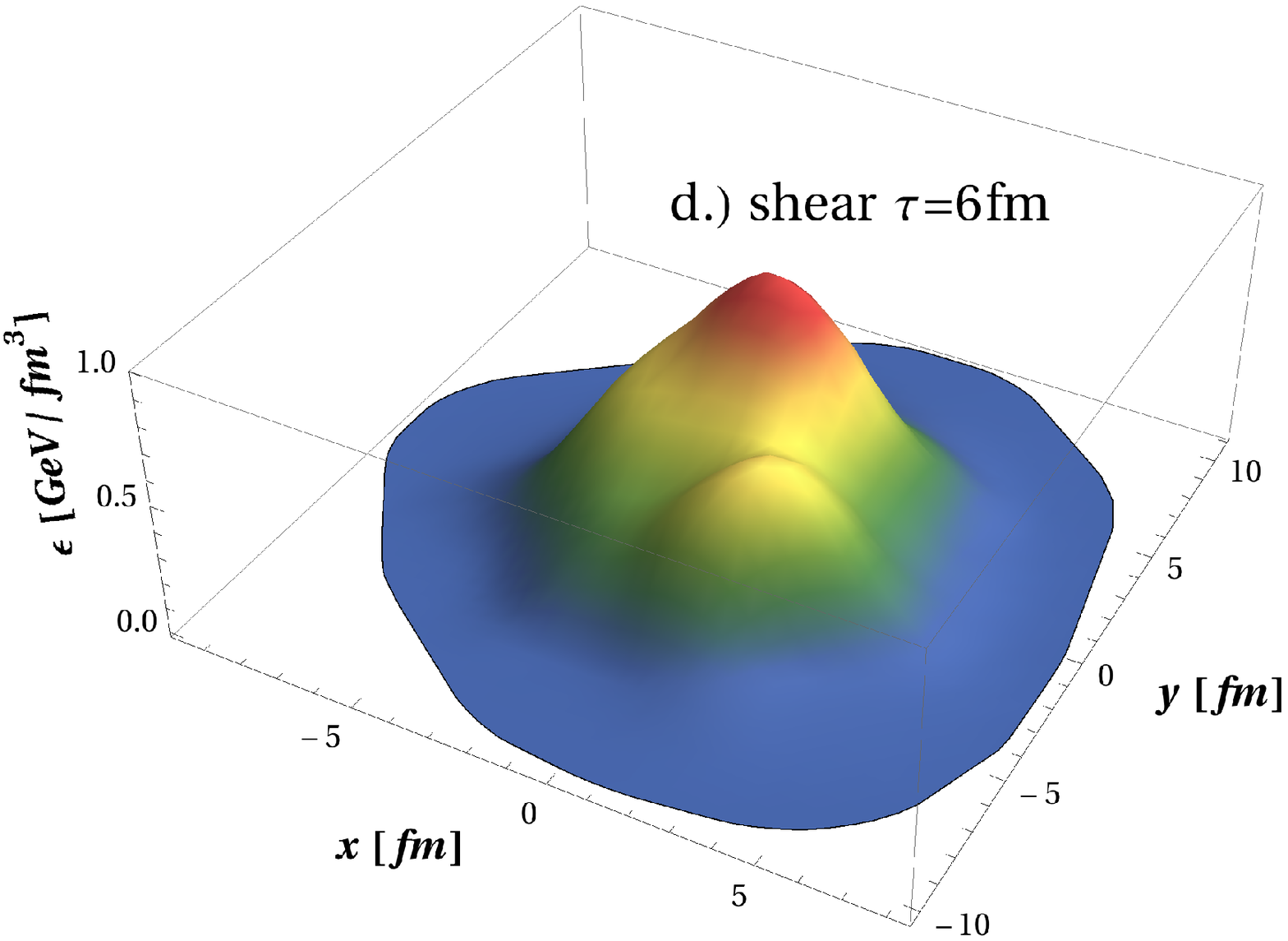} &
\includegraphics[width=0.5\textwidth]{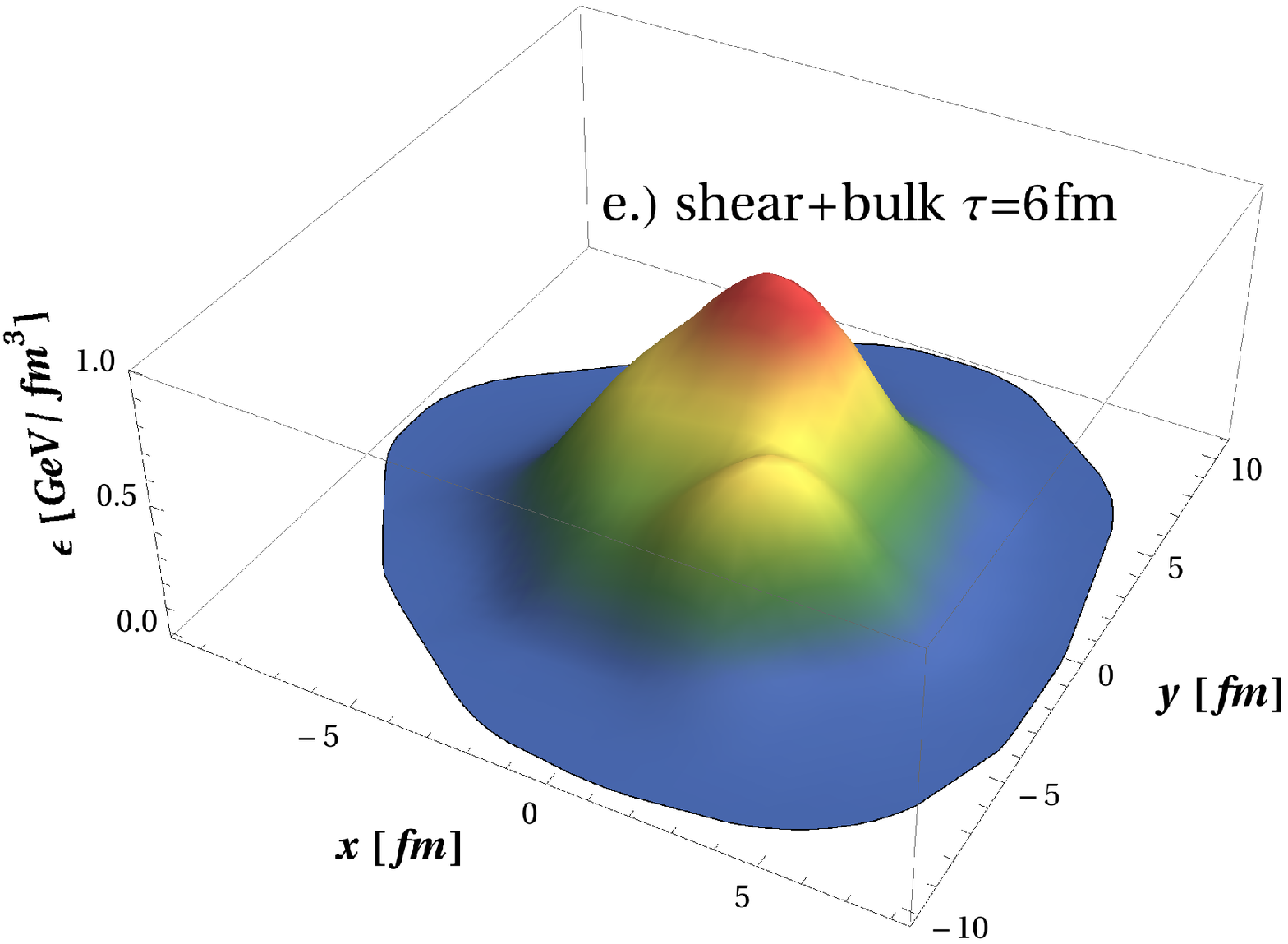} \\
\end{tabular}%
\caption{(Color online) Energy Density distribution for a random event in peripheral ($20-30\%$) collisions at RHIC for different times. a.) initial time $\tau_0=1$ fm. Energy distribution at $\tau=6$ fm for ideal hydrodynamics in b.), c.) viscous hydrodynamics with only bulk viscosity, d.) viscous hydrodynamics with only shear viscosity, and e.) viscous hydrodynamics with both bulk and shear viscosity effects. The transport coefficients are the ones shown in Fig.\ \ref{fig:transco}.}
\label{tab:eden}
\end{figure}

In this section we describe the interplay between shear and bulk viscosity within the fluid evolution. To do so we consider here first the effects of viscosity on the energy density over an interval of $\Delta\tau=5$ fm (we start the evolution at $\tau_0=1$ fm and plot the energy density profile at $\tau=6$ fm) for a random initial condition shown in Fig.\ \ref{tab:eden} a.) for RHIC's $20-30\%$ centrality class.  We then include plots of the fluid evolution of the energy density at $\tau=6$ fm for b.) ideal hydrodynamics, c.) viscous hydrodynamics with only bulk viscosity, d.) viscous hydrodynamics with only shear viscosity effects, and e.) viscous hydrodynamics with both bulk and shear viscosity. 

One can clearly see in Fig.\ \ref{tab:eden} that there are qualitative differences between the ideal and viscous fluids.  The ideal hydrodynamical evolution preserves most of the initial structure of the energy density even after $\Delta\tau=5$ fm.  The bulk viscous case evolution does not maintain as many peaks and valleys at the ideal case but still displays more structure than both the shear and shear+bulk hydro events.  Moreover, it is flatter than all the other events (recall that bulk viscosity acts against radial expansion).  For the energy density profile we are able to see no difference between the bulk and the shear+bulk case, which indicates that the shear viscosity dominates the viscous corrections to energy density throughout the hydrodynamical evolution. This is not surprising considering that our chosen $\zeta/s$ is relatively small in comparison to $\eta/s$ and the energy density is a relatively robust observable.

\begin{table}[htbp]
\begin{center}
\begin{tabular}{|c|c|c|c|c|c|c|}
\hline
 & $\langle \Pi\rangle$ & $\sigma^2_{\Pi}$ & $\langle \Pi\rangle_{early}$ & $(\sigma^2_{\Pi})_{early}$ & $\langle \Pi\rangle_{late}$ & $(\sigma^2_{\Pi})_{late}$ \\ \hline
\hline
0-10\% & 1.79\% & 8.59\% & 1.14\% & -59.72\% & 2.03\% & 20.50\% \\ \hline
10-20\% & 2.48\% & 8.95\% & 2.89\% & -52.37\% & 2.19\% & 20.59\% \\ \hline
20-30\% & 2.87\% & 8.96\% & 4.07\% & -40.70\% & 2.02\% & 20.66\% \\ \hline
30-40\% & 3.49\% & 9.15\% & 3.47\% & -36.96\% & 2.15\% & 19.97\% \\ \hline
40-50\% & 4.14\% & 9.11\% & 3.52\% & -37.23\% & 2.00\% & 20.86\% \\ \hline
50-60\% & 4.98\% & 9.23\% & 6.27\% & -22.55\% & 2.28\% & 19.73\% \\ \hline
\end{tabular}
\end{center}
\caption{Percentage change of the mean values of the bulk pressure $\Pi$ and its corresponding variance $\sigma^2_{\Pi}$ averaged over all events for different centrality classes due to the presence of shear viscosity. $\langle \Pi\rangle$ and $\sigma^2_{\Pi}$ takes into account the parts of the fluid that have frozen out throughout the whole time evolution, $\langle \Pi\rangle_{early}$ and $(\sigma^2_{\Pi})_{early}$ are computed using only the parts of the fluid that have frozen out between $\tau_0=1$ fm and $\tau=2$ fm, $\langle \Pi\rangle_{late}$ and $(\sigma^2_{\Pi})_{late}$ are computed using only the parts of the fluid that have frozen out in the last fm of the time evolution.}
\label{percchangebulk}
\end{table}
While there is little difference between the case involving only shear and shear+bulk within the energy density profile, it is interesting to see if  an effect shows up in the nonzero components of the shear stress tensor $\pi^{\mu\nu}$ and, additionally, in the bulk pressure $\Pi$. 

To see how the inclusion of shear viscosity affects the bulk pressure, we first look at the mean (averaged over all the SPH particles) of the bulk pressure for each individual event, $\left(\Pi\right)_{ev}$, and its corresponding variance and define 
\begin{eqnarray}
(\Pi)_{ev}&=&100\,\frac{\left(\Pi_{sb}\right)_{ev}-\left(\Pi_{b}\right)_{ev}}{\left(\Pi_{b}\right)_{ev}}\nonumber\\
(\sigma^2_{\Pi})_{ev}&=&100\,\frac{\left(\sigma^2_{\Pi_{sb}}\right)_{ev}-\left(\sigma^2_{\Pi_{b}}\right)_{ev}}{\left(\sigma^2_{\Pi_{b}}\right)_{ev}},
\end{eqnarray}\label{eqn:perchange}
where $\left(\Pi_{sb}\right)_{ev}$ is the mean bulk pressure of a given event $ev$ with the corresponding variance $\left(\sigma^2_{\Pi_{sb}}\right)_{ev}$ when the equations of motion include both shear viscosity and bulk viscosity while $\left(\Pi_{b}\right)_{ev}$ is the mean bulk pressure of the same event $ev$ with the corresponding variance $\left(\sigma^2_{\Pi_{b}}\right)_{ev}$ when the equations of motion include only bulk viscosity. We then average the percentage change over all the events within each individual centrality class such that we look at the mean percentage change of the bulk pressure $\langle\Pi\rangle$ and the mean percentage change of the variance of the bulk pressure $\langle\sigma^2_{\Pi}\rangle$ over all the events. 

In Table \ref{percchangebulk}, $\langle \Pi\rangle$ and $\sigma^2_{\Pi}$ takes into account the parts of the fluid that have frozen out throughout the whole time evolution, $\langle \Pi\rangle_{early}$ and $(\sigma^2_{\Pi})_{early}$ are computed using only the parts of the fluid that have frozen out between $\tau_0=1$ fm and $\tau=2$ fm, $\langle \Pi\rangle_{late}$ and $(\sigma^2_{\Pi})_{late}$ are computed using only the parts of the fluid that have frozen out in the last fm of the time evolution. In general, we see that when shear viscosity is added to our hydrodynamical evolution the changes in the bulk pressure are not large.  The mean percentage change of the bulk pressure $\langle\Pi\rangle$ is small and only increases for more peripheral events.  Also, the mean percentage change of the variance of the bulk pressure $\langle\sigma^2_{\Pi}\rangle$ is around $9\%$ across all centrality classes, which means that the shear viscosity slight increases $\Pi$  and also makes the distribution of $\Pi$ only $9\%$ wider on average.  The mean percentage change of $\langle \Pi\rangle_{early}$ and $\langle \Pi\rangle_{late}$ are positive and $<10\%$ and one can see that the percentage change of the variance $(\sigma^2_{\Pi})$ decreases significantly at early times while at late times it increases by $\sim 20\%$ for all centrality classes due to the inclusion of shear. This shows that even though the mean bulk pressure is not that affected by the presence of shear, its distribution computed event by event becomes sharper around the mean at early times and gets broadened at late times.

\begin{table}
\begin{center}
\begin{tabular}{|l|c|c|c|c|}
\hline
Centrality & $\langle \pi^{00}\rangle$ & $\sigma^2_{\pi^{00}}$ & $\langle \pi^{12}\rangle$ & $\sigma^2_{\pi^{12}}$ \\ 
 \hline
0-10\% & -17.61\% & -19.09\% & -2.87\% & -8.50\%  \\ \hline
10-20\% & -17.77\% & -18.53\% & -2.25\% & -8.45\% \\ \hline
20-30\% & -19.22\% & -18.56\% &-3.48\% & -8.44\% \\ \hline
30-40\% & -22.98\% & -18.53\% &-3.26\% & -8.35\% \\ \hline
40-50\% & -38.11\% & -19.37\% & -2.81\% & -8.01\% \\ \hline
50-60\% & -44.63\% & -19.61\% & -5.05\% & -7.68\% \\ \hline
\end{tabular}
\end{center}
\caption{The percentage change in the mean values and variance of the $\pi^{00}$ and $\pi^{12}$ components of the shear stress tensor $\pi^{\mu\nu}$ averaged over all events and all SPH particles due to the inclusion of bulk viscosity in the time evolution. These quantities are computed taking into account the parts of the fluid that have frozen out throughout the whole time evolution.}
\label{perchangshear}
\end{table}
While the effects of shear on the bulk pressure are not large, the effects of bulk viscosity on the shear stress tensor are not so trivial. In Table \ref{perchangshear} we show the percentage change in the mean values and variance of the $\pi^{00}$ and $\pi^{12}$ components of the shear stress tensor $\pi^{\mu\nu}$ averaged over all events and all SPH particles due to the inclusion of bulk viscosity in the time evolution. These quantities are computed taking into account the parts of the fluid that have frozen out throughout the whole time evolution. We note that since $\pi^{\mu\nu}$ is traceless, $\pi^{00} = \pi^{11}+\pi^{22}+\tau^2 \pi^{33}$. One can see that the inclusion of bulk viscosity considerably affects $\langle \pi^{00}\rangle$: there is a suppression in its average value that increases towards more peripheral collisions while its variance also decreases by $\sim 20\%$ for all centralities due to the nonzero bulk viscosity. Therefore, the distribution of $\pi^{00}$ has a smaller mean and becomes sharper around the mean due to bulk viscosity. This occurs because bulk viscosity dampens out radial disturbances of pressure and flow, which in turn should decrease the diagonal components of the shear stress tensor. On the other hand, $\pi^{12}$ is only slightly affected by bulk viscosity both in terms of its mean and variance, which could be expected from symmetry arguments.

In Table \ref{perchangshearearly} we show the corresponding quantities obtained from the parts of the fluid that have already frozen out after 1 fm passed the initial time $\tau_0$. In this case, one can see that at early times the modification of the fluid velocity due to bulk viscosity has not yet affected the shear stress tensor by much. The mean value of $\pi^{00}$ decreases by $< 7\%$ in a way that is almost independent on centrality. This should be contrasted to the results in Table \ref{perchangshear} which took into account the modification in this component throughout the whole time evolution due to bulk viscosity, which becomes more significant in peripheral collisions. Its variance decreases by $\sim 13\%$ in the most central collisions while it for peripheral collisions the suppression is $\sim 15\%$. Once more, the distribution of $\pi^{12}$ is only slightly affected ($< 10\%$) by the presence of bulk viscosity.

\begin{table}
\begin{center}
\begin{tabular}{|l|c|c|c|c|}
\hline
Centrality & $\langle \pi^{00}\rangle_{early}$ & $(\sigma^2_{\pi^{00}})_{early}$ & $\langle \pi^{12}\rangle_{early}$ & $(\sigma^2_{\pi^{12}})_{early}$ \\ 
  \hline
0-10\% & -6.66\% & -12.79\% & -5.94\% & -10.66\% \\ \hline
10-20\% & -5.32\% & -11.31\% & -4.87\% & -9.46\% \\ \hline
20-30\% & -6.07\% & -12.72\% & -4.81\% & -9.15\% \\ \hline
30-40\% & -7.01\% & -14.08\% &-4.80\% & -9.19\% \\ \hline
40-50\% & -4.75\% & -9.00\% & -4.75\% & -8.99\% \\ \hline
50-60\% & -6.83\% & -15.02\% &-4.63\% & -8.76\% \\ \hline
\end{tabular}
\end{center}
\caption{The percentage change in the mean values and variance of the $\pi^{00}$ and $\pi^{12}$ components of the shear stress tensor $\pi^{\mu\nu}$ averaged over all events and all SPH particles due to the inclusion of bulk viscosity in the time evolution. These quantities are computed taking into account only the parts of the fluid that have already frozen for early times (between $\tau=\tau_0$ and $\tau=2$ fm). }
\label{perchangshearearly}
\end{table}

\begin{table}
\begin{center}
\begin{tabular}{|l|c|c|c|c|}
\hline
Centrality & $\langle \pi^{00}\rangle_{late}$ & $(\sigma^2_{\pi^{00}})_{late}$ & $\langle \pi^{12}\rangle_{late}$ & $(\sigma^2_{\pi^{12}})_{late}$ \\ 
 \hline
0-10\% & -17.68\% & -29.13\% &-5.94\% & -10.80\% \\ \hline
10-20\% & -15.98\% & -29.09\% & -4.80\% & -9.38\% \\ \hline
20-30\% & -15.45\% & -28.56\% & -4.77\% & -9.06\% \\ \hline
30-40\% & -14.97\% & -28.28\% & -4.88\% & -9.34\% \\ \hline
40-50\% & -13.83\% & -27.91\% & -4.80\% & -9.20\% \\ \hline
50-60\% & -12.75\% & -26.18\% &  -4.50\% & -8.51\% \\ \hline
\end{tabular}
\end{center}
\caption{The percentage change in the mean values and variance of the $\pi^{00}$ and $\pi^{12}$ components of the shear stress tensor $\pi^{\mu\nu}$ averaged over all events and all SPH particles due to the inclusion of bulk viscosity in the time evolution. These quantities are computed taking into account only the parts of the fluid that have frozen during the last fm of the time evolution.}
\label{perchangshearlate}
\end{table}

At later times we see a larger effect on the shear stress tensor components from the bulk viscosity. In Table \ref{perchangshearlate} we see that for almost every case both the mean and variance, regardless of centrality class, are significantly larger for late freeze-out (the last $\Delta \tau=1fm$ of the hydrodynamical evolution).  This indicates that as the hydrodynamical evolution progresses the effects of bulk viscosity are more visible, which in the end is consistent with the results in Table \ref{perchangshear}. This occurs because it takes some time for the bulk viscosity to affect the flow and then the shear tensor and, consequently, the shear stress tensor. Furthermore, one expects that by lowering the freeze-out temperature (here we use $T_{0}=150$ MeV) one can increase the effects from bulk viscosity.  By the same reason, going from RHIC to LHC energies one would expect that bulk viscosity becomes more relevant to the dynamical evolution of the system since at large energies the fluid stays in the QGP phase for a longer period of time.

Tables \ref{percchangebulk}-\ref{perchangshearlate} suggest that the interplay between bulk and shear viscosities have a very non-trivial, non-linear effect on the shear stress tensor and bulk pressure already during the hydrodynamical evolution itself. While the shear only slightly increases the mean value of $\Pi$, the inclusion of bulk viscosity leads to a significant suppression of the shear stress tensor components. This indicates that the expected suppression of flow harmonics due to shear viscosity can be softened by the presence of bulk viscosity. In fact, in our previous work \cite{Noronha-Hostler:2013gga} we suggested that it may be possible for the bulk viscosity-driven enhancement of the integrated flow harmonics $vn$'s compensate for the expected damping of these coefficients due to shear viscosity.  However, it appears that their relationship is more complicated than we initially believed. 


\section{Results for the Flow Harmonics}\label{sec:results}

In this section we show the results for both the $p_T$-integrated and the differential flow harmonics taking into account the effect of bulk and shear viscosities. We use the event plane method \cite{Poskanzer:1998yz} to calculate the event plane angles $\psi_n$'s and a detailed explanation of the method as done in v-USPhydro can be found in \cite{Noronha-Hostler:2013gga}. Additionally, in each centrality class we consider 150 events on an event-by-event basis (we have checked that the results found here are robust with respect to the inclusion of more events). 

Unless stated otherwise, for the $p_T$-integrated $v_n$'s we take the limits of integration to be $p_T=0-5$ GeV.  However, due to issues previously discussed with the bulk viscous corrections within Cooper-Frye freeze-out \cite{Noronha-Hostler:2013gga}, the overall viscous correction to the particle distribution at freeze-out can become larger (and negative) than the equilibrium contribution at intermediate values of $p_T$ (which would lead to a negative particle spectra for those values of $p_T$) if the viscous transport coefficients are large (for the coefficients shown in Fig.\ \ref{fig:transco} this problem does not occur). In order to avoid such problems in the spectra and the integrated $v_n$'s when $\zeta/s$ is 10 times larger than that in Eq.\ (\ref{eqn:adszeta}), we did not take into account the negative contribution from the corresponding part of the integral over $p_T$.

\subsection{Integrated $v_n$'s}\label{sec:intvns}


\begin{figure}[ht]
\centering
\begin{tabular}{cc}
\includegraphics[width=0.5\textwidth]{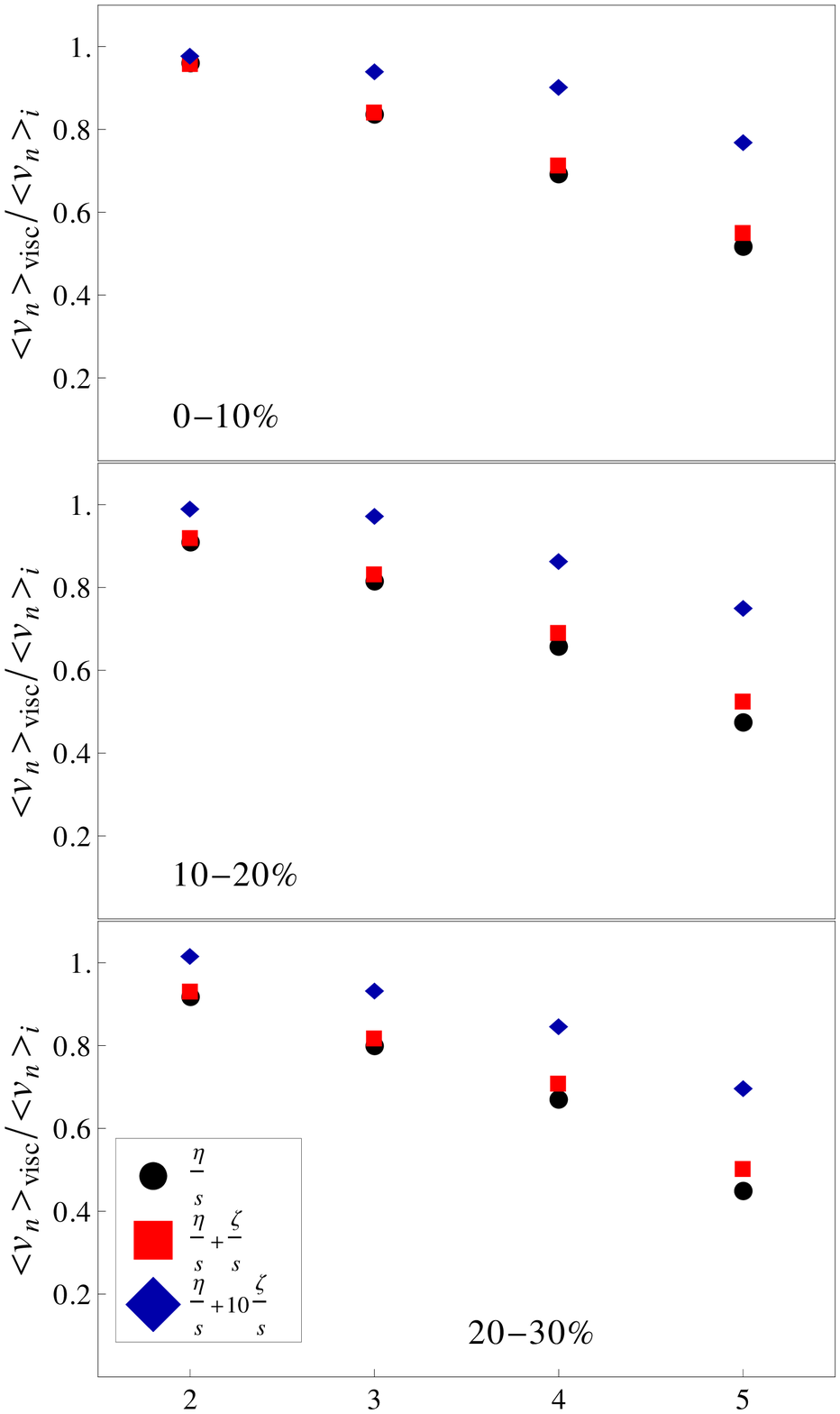} & %
\includegraphics[width=0.5\textwidth]{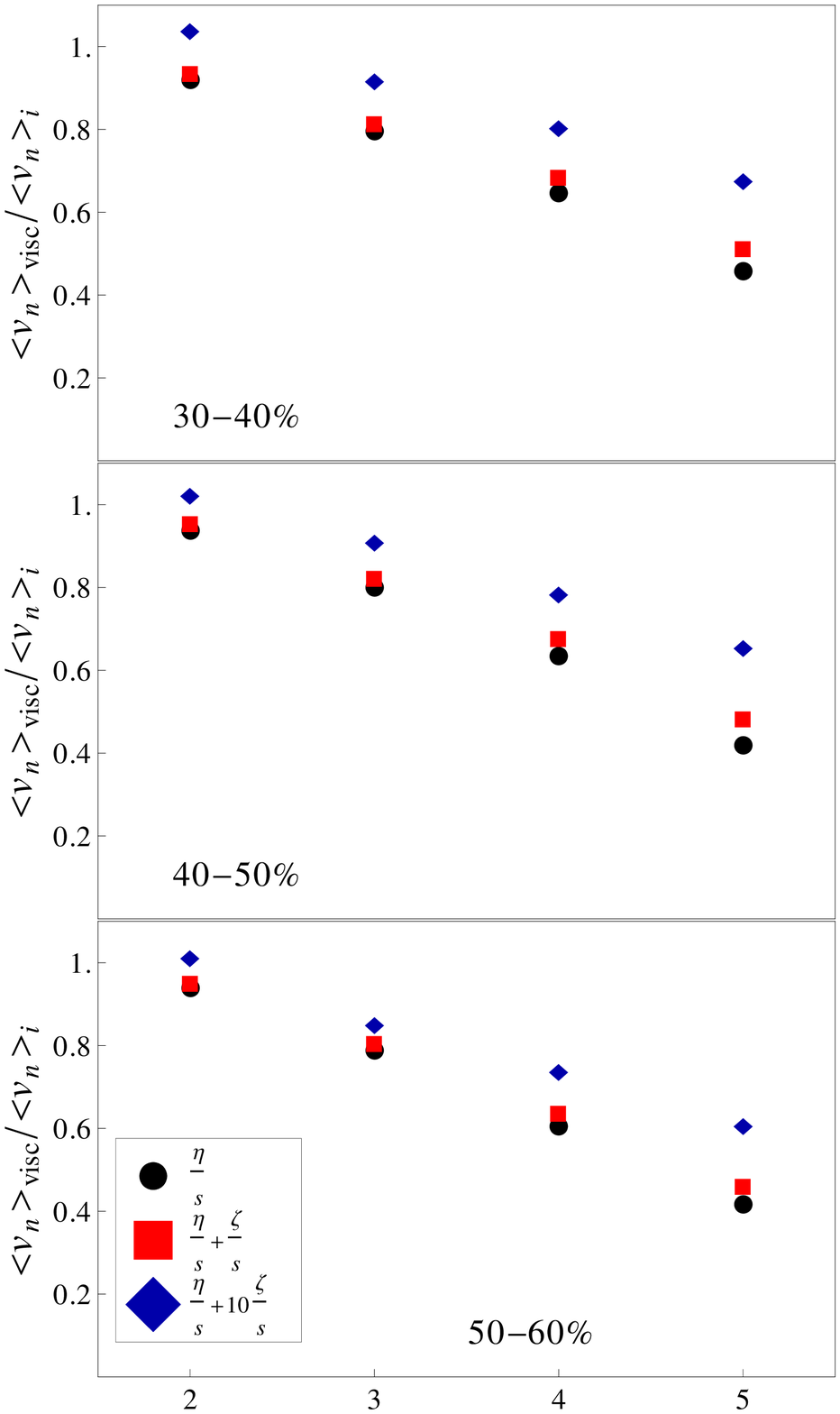} \\ 
\end{tabular}%
\caption{(Color online) Ratio between the integrated $v_n$'s of viscous and ideal hydrodynamics of direct pions over all centralities at RHIC computed using the Moments Method (MOM) for the bulk viscosity contribution at freeze-out. The circles correspond to the case where only shear viscosity is taken into account, the squares represent the case with both shear and bulk viscosity, while the diamonds correspond to the case where shear and bulk are included but $\zeta/s$ is multiplied by a factor of 10.}
\label{tab:vintmom}
\end{figure}

In Fig.\ \ref{tab:vintmom} we show the ratio between the integrated $v_n$'s of direct pions at RHIC of viscous and ideal hydrodynamics across all centrality classes investigated in this paper. These quantities were computed using the moments method to determine the viscous correction to the particle distribution at freeze-out. The transport coefficients we used are defined in Section \ref{sec:bulkshear}. 
The $v_n$ dependence on $n$ has the steepest curve when only shear viscosity is included.  The effect of bulk viscosity combined with shear viscosity slightly increases the $v_n$'s, in accordance with the conclusions found in \cite{Noronha-Hostler:2013gga} that the bulk viscosity slightly increases the $v_n$'s.  It is, however, a small increase since our chosen $\zeta/s$ is significantly smaller than $\eta/s$. The suppression of shear viscosity effects here occurs because, as shown in the previous section, the inclusion of bulk viscosity decreases the magnitude of the shear stress tensor components.

Finally, when one includes the effect of a ``large" bulk viscosity, i.e., $10\zeta/s$ we see that the $v_n$'s are shifted upwards much closer to the ideal case.  In this case the bulk viscosity-driven suppression of the shear stress tensor is very significant. We note here that our initial bulk viscosity is so small that even after multiplying by a factor of 10, its peak is still not as large as the minimum of the shear viscosity (see Fig.\ \ref{fig:transco}).  Thus, we do not expect that the bulk viscosity can completely counteract shear viscous effects even in this case. Furthermore, due to above mentioned limitations in the $\delta f$, for the case with $10\zeta/s$ one can only integrate to $p_T=0.8$ GeV (which in any case is the integration interval that gives the major contribution to this quantity) before the spectrum becomes negative.

We note that the inclusion of bulk viscosity has the net effect to decrease the difference between $v_2$ and $v_3$ in central collisions. In fact, in the case of $10\zeta/s$ there is very little difference between $v_2$ and $v_3$ in the most central collisions, which is not the case towards peripheral collisions. If the bulk viscosity of the QGP is not really much smaller than $\eta/s$, further improvements to the viscous correction $\delta f$ involving bulk and shear are necessary for a more accurate calculation of $v_n$'s to verify the trend regarding $v_2$ and $v_3$ in central collisions found here.

\begin{figure}[ht]
\centering
\begin{tabular}{cc}
\includegraphics[width=0.5\textwidth]{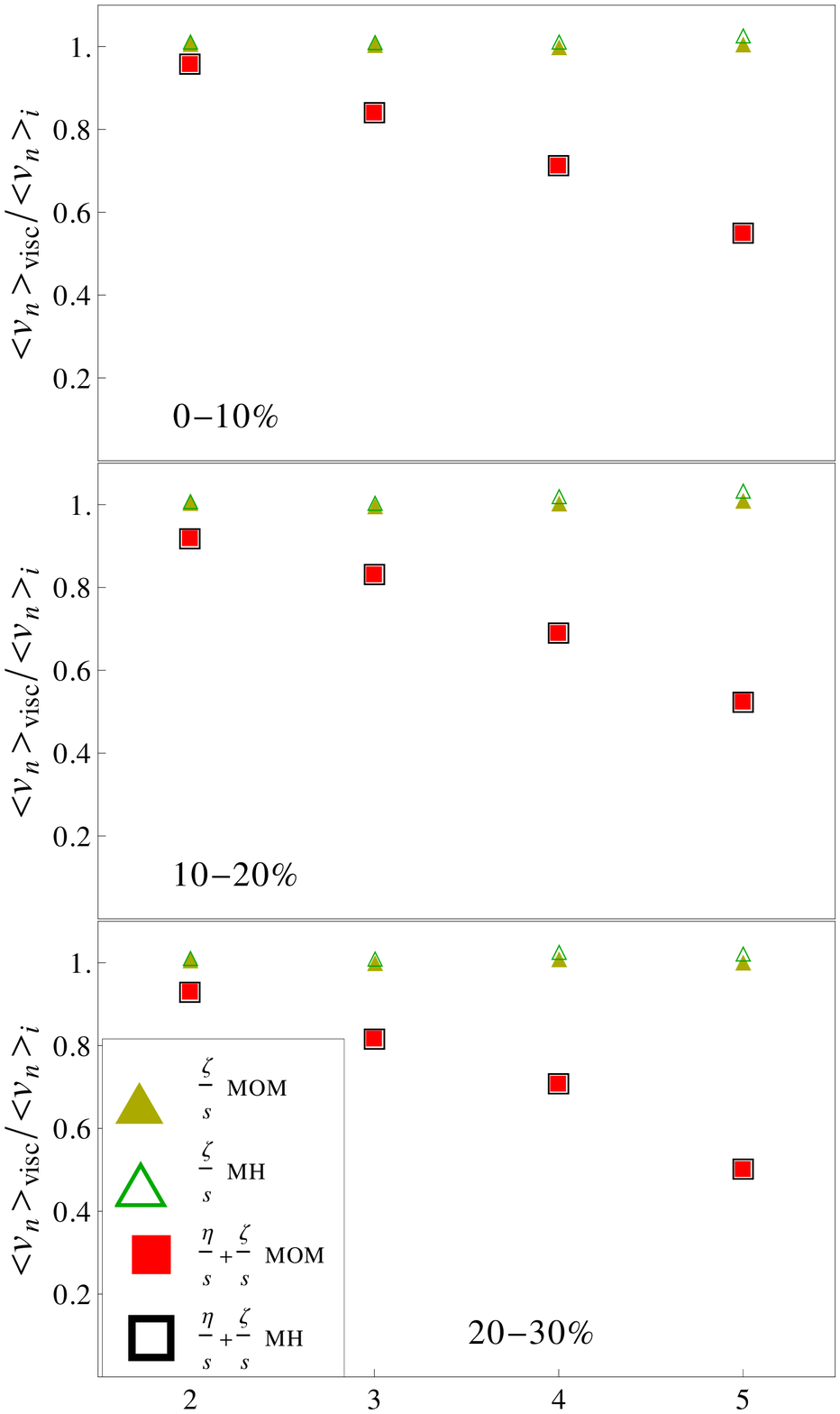} & %
\includegraphics[width=0.5\textwidth]{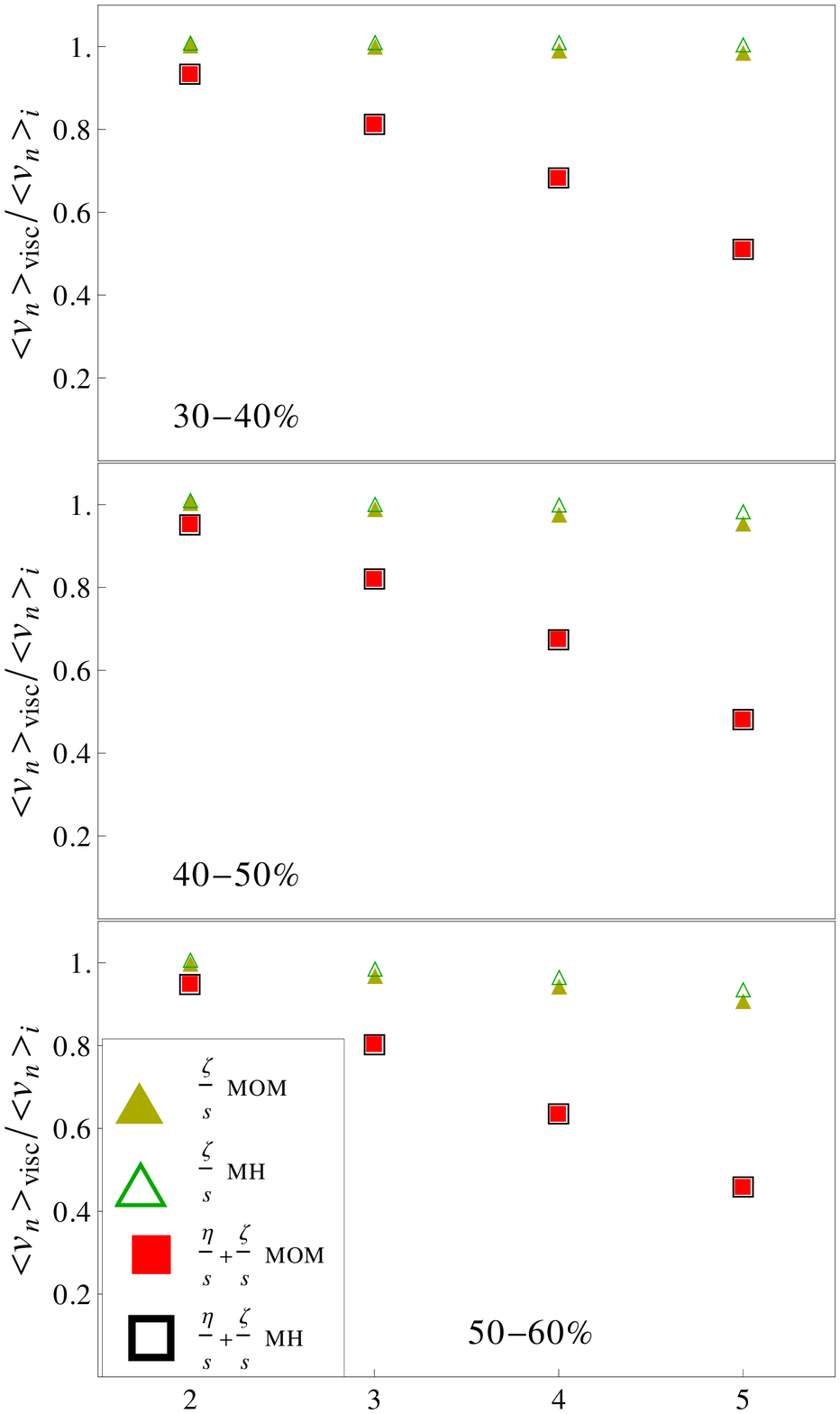} \\ 
\end{tabular}%
\caption{(Color online) Ratio between the integrated $v_n$'s of viscous and ideal hydrodynamics of direct pions over all centralities at RHIC computed using the Moments Method (MOM) and the Monnai-Hirano (MH) formulas for the bulk viscosity contribution at freeze-out. The filled triangles correspond to the case with only bulk viscosity with $\delta f^{Bulk}$ from MOM while the empty triangles correspond to the analogous case computed with $\delta f^{Bulk}$ from MH. The solid squares correspond to the case with shear and bulk viscosities with $\delta f^{Bulk}$ from MOM while empty squares correspond to the analogous case computed with $\delta f^{Bulk}$ from MH.}
\label{tab:vintdfcomp}
\end{figure}
In Fig.\ \ref{tab:vintdfcomp} we compare the difference between two different choices of $\delta f^{Bulk}$ viscous corrections within Cooper-Frye freeze-out.  As discussed in the previous section, the moments method is derived in \cite%
{Denicol:2012cn,Denicol:2012yr} and while MH comes from \cite{Monnai:2009ad}. It appears that at least in the case of the integrated $v_n$'s there is almost no difference between the two methods for all centrality classes and $v_n$'s.  The moments method shows a slight increase for the integrated $v_n$'s for both bulk and shear+bulk but the change is very small. This is because the integrated $v_n$'s are more dependent on the lower $p_T$ region wherein there is little difference between MOM and MH.

\subsection{Differential $v_n$'s}\label{eqn:difvns}

\begin{figure}[ht]
\centering
\begin{tabular}{cc}
\includegraphics[width=0.5\textwidth]{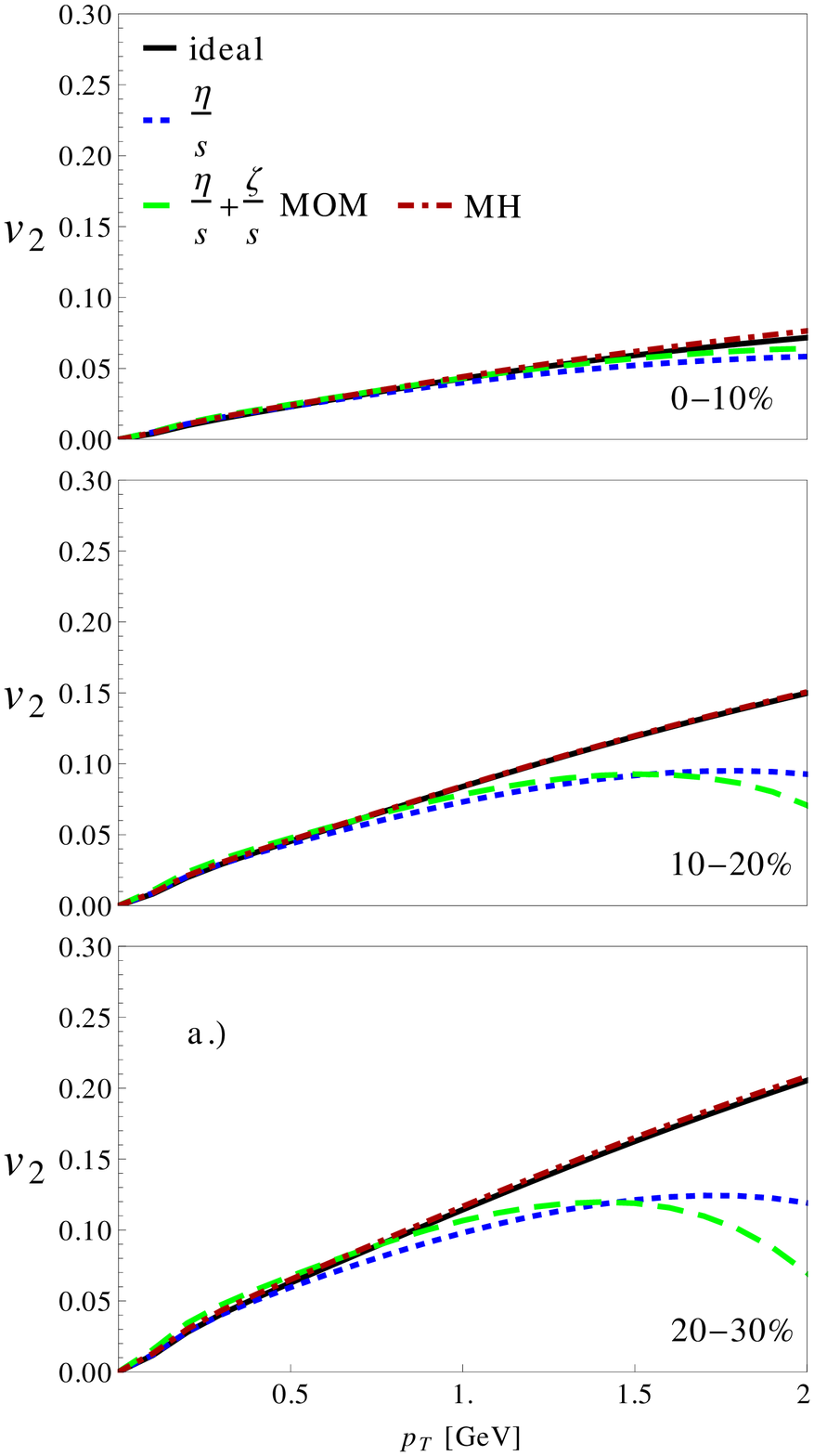} & %
\includegraphics[width=0.5\textwidth]{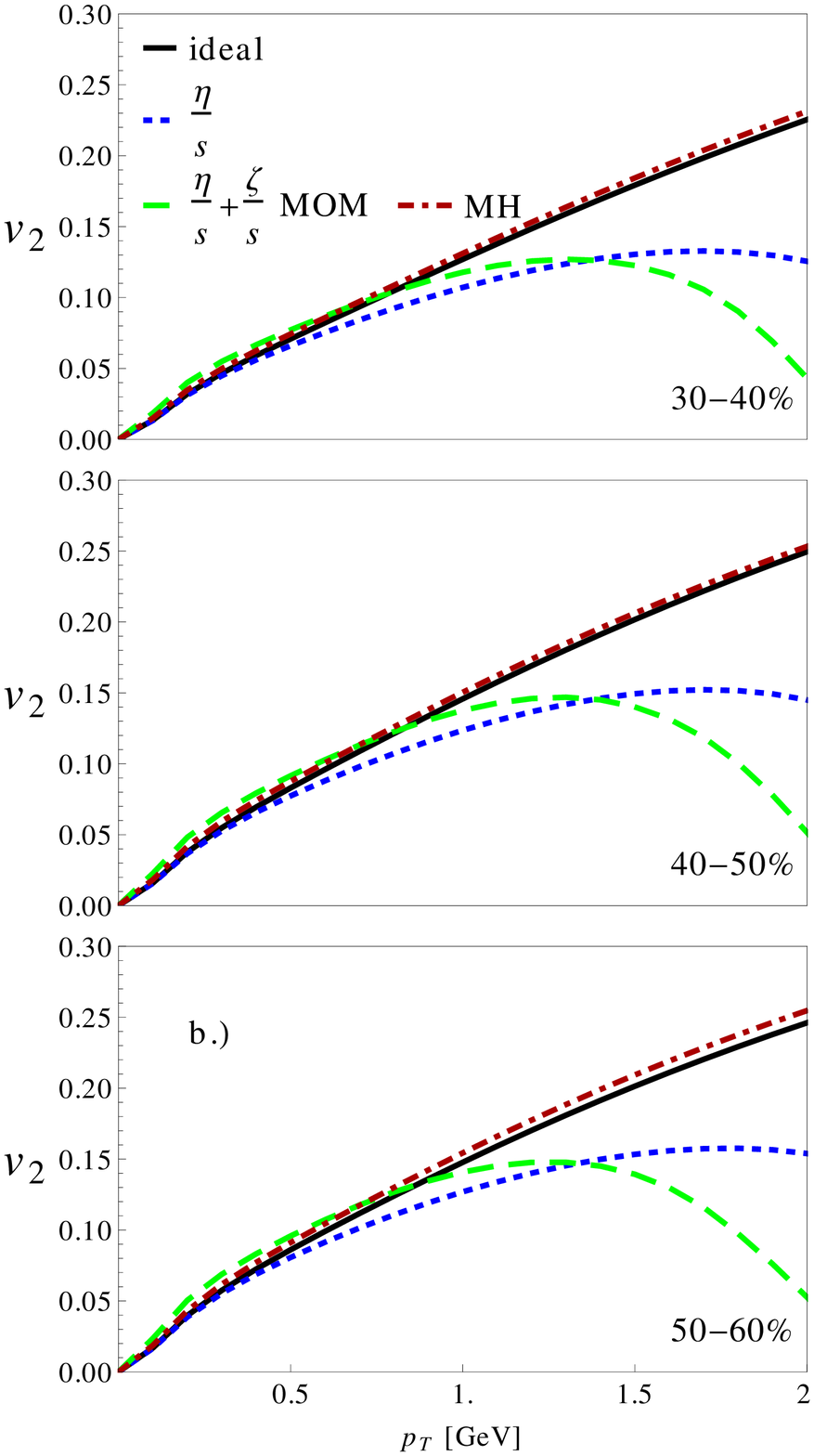} \\ 
\end{tabular}%
\caption{(Color online) Comparison of $v_2$ of direct pions across the centrality classes $0-10\%$, $10-20\%$, $20-30\%$ (in column a.), $30-40\%$, $40-50\%$, and $50-60\%$ (in column b.) for RHIC. The solid black line denotes the ideal hydro result, the short-dashed blue line was computed taking into account only shear viscosity, the long-dashed green curve was computed using shear+bulk with the moments method expression for the freeze-out while the dark red dotted-dashed curve corresponds to shear+bulk with the MH formula.}
\label{tab:v2}
\end{figure}

\begin{figure}[ht]
\centering
\begin{tabular}{cc}
\includegraphics[width=0.5\textwidth]{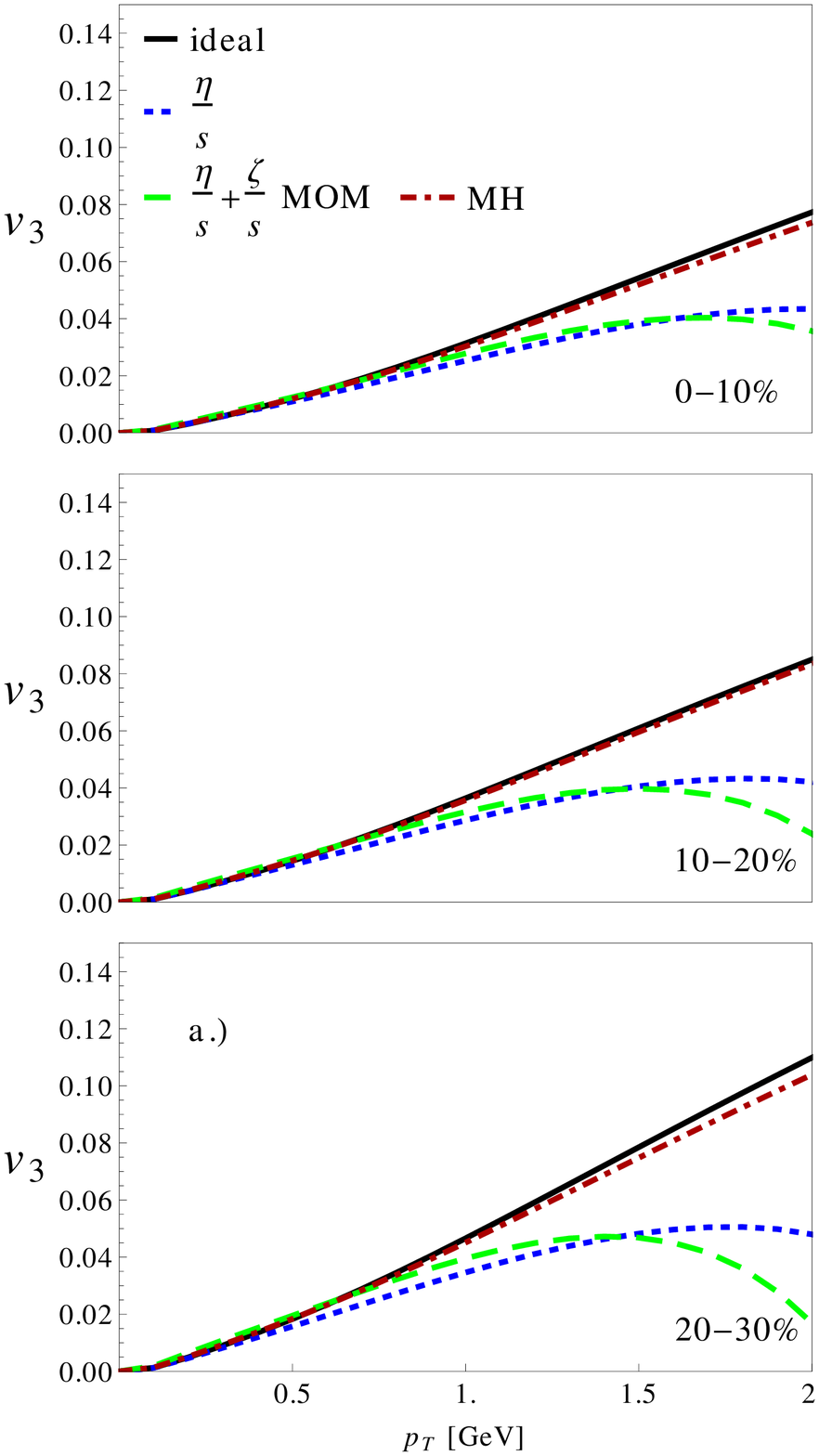} & %
\includegraphics[width=0.5\textwidth]{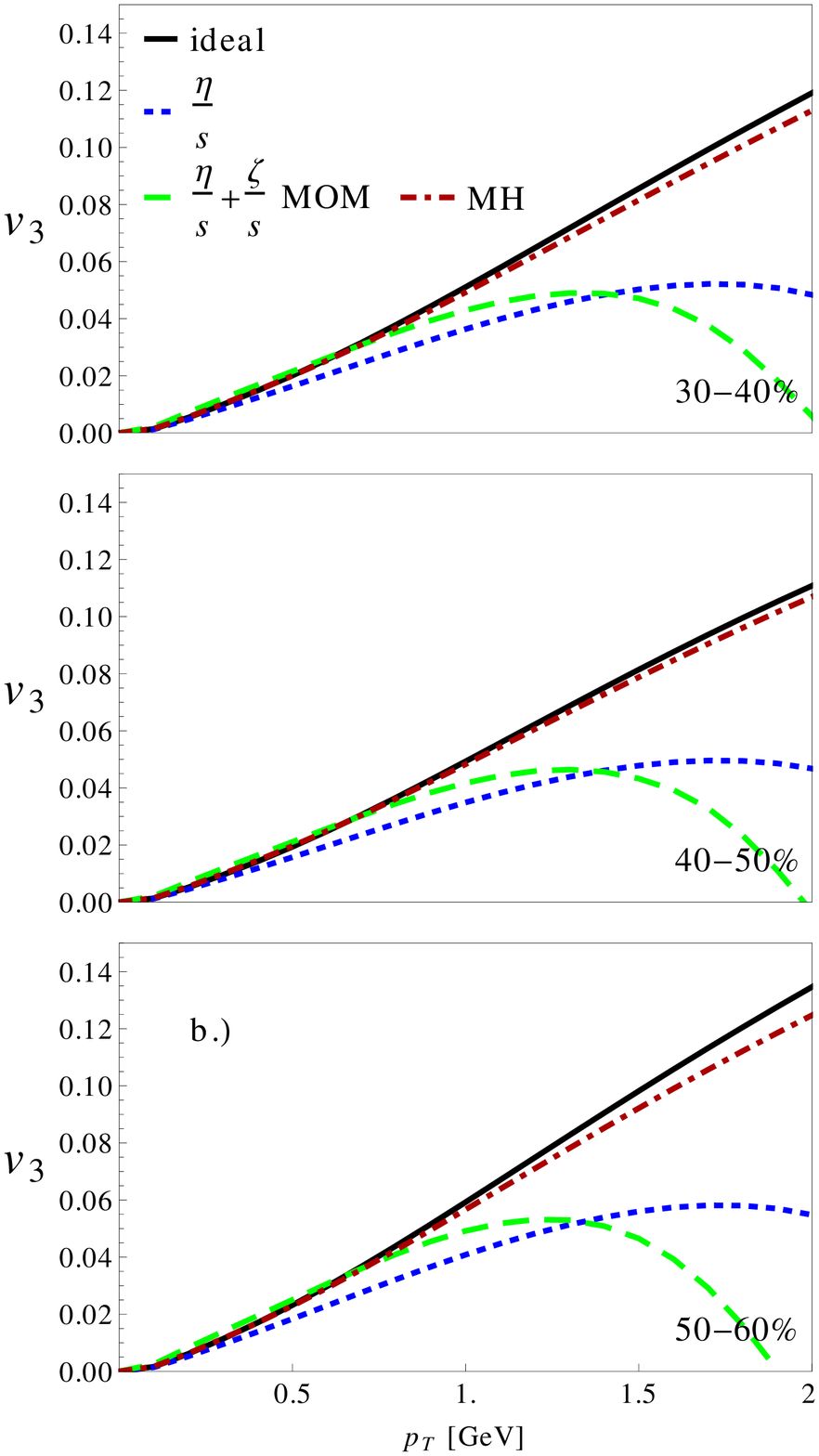} \\ 
\end{tabular}%
\caption{(Color online) Comparison of $v_3$ of direct pions across the centrality classes $0-10\%$, $10-20\%$, $20-30\%$ (in column a.), $30-40\%$, $40-50\%$, and $50-60\%$ (in column b.) for RHIC. The solid black line denotes the ideal hydro result, the short-dashed blue line was computed taking into account only shear viscosity, the long-dashed green curve was computed using shear+bulk with the moments method expression for the freeze-out while the dark red dotted-dashed curve corresponds to shear+bulk with the MH formula.}
\label{tab:v3}
\end{figure}

\begin{figure}[ht]
\centering
\begin{tabular}{cc}
\includegraphics[width=0.5\textwidth]{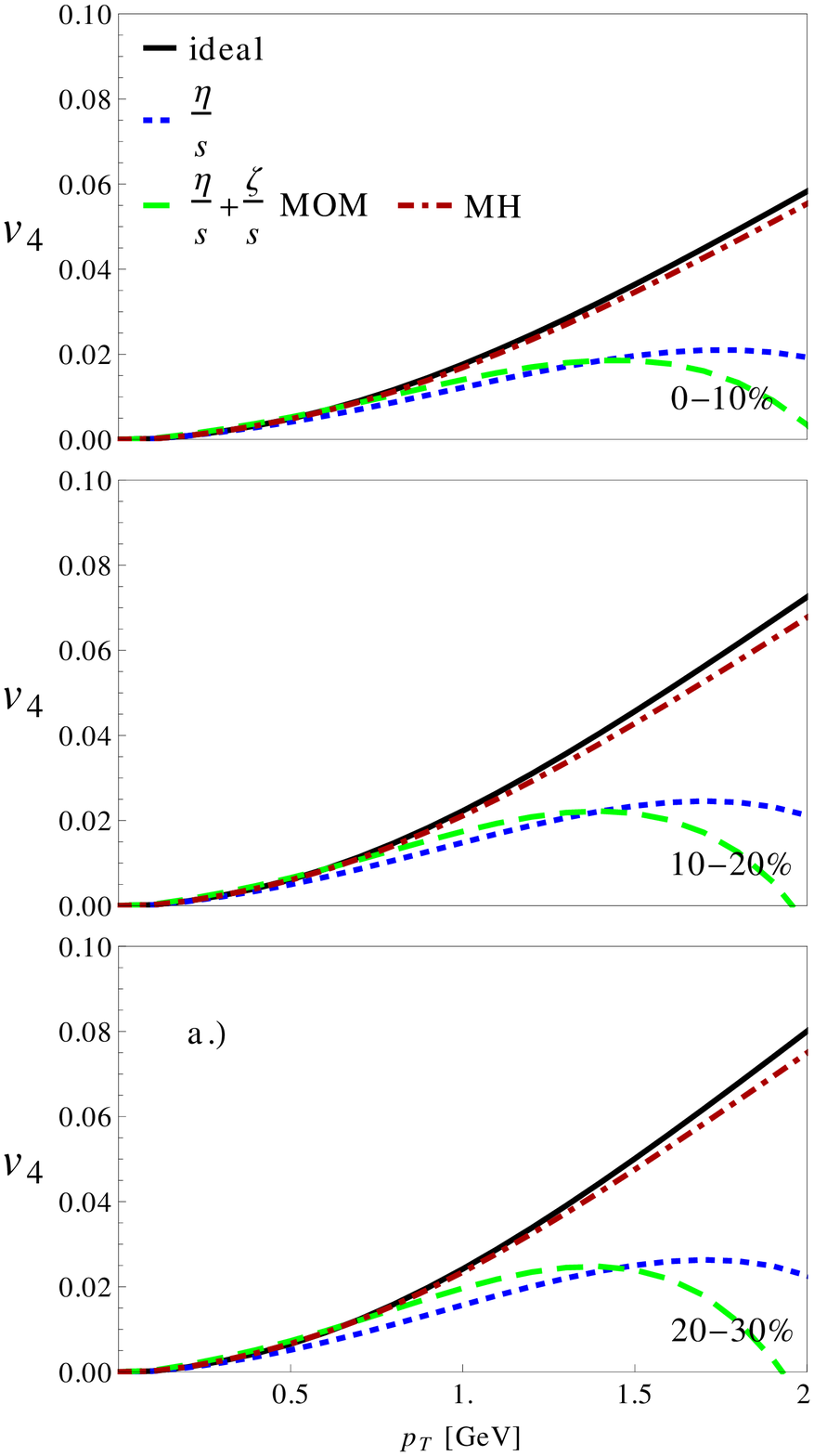} & %
\includegraphics[width=0.5\textwidth]{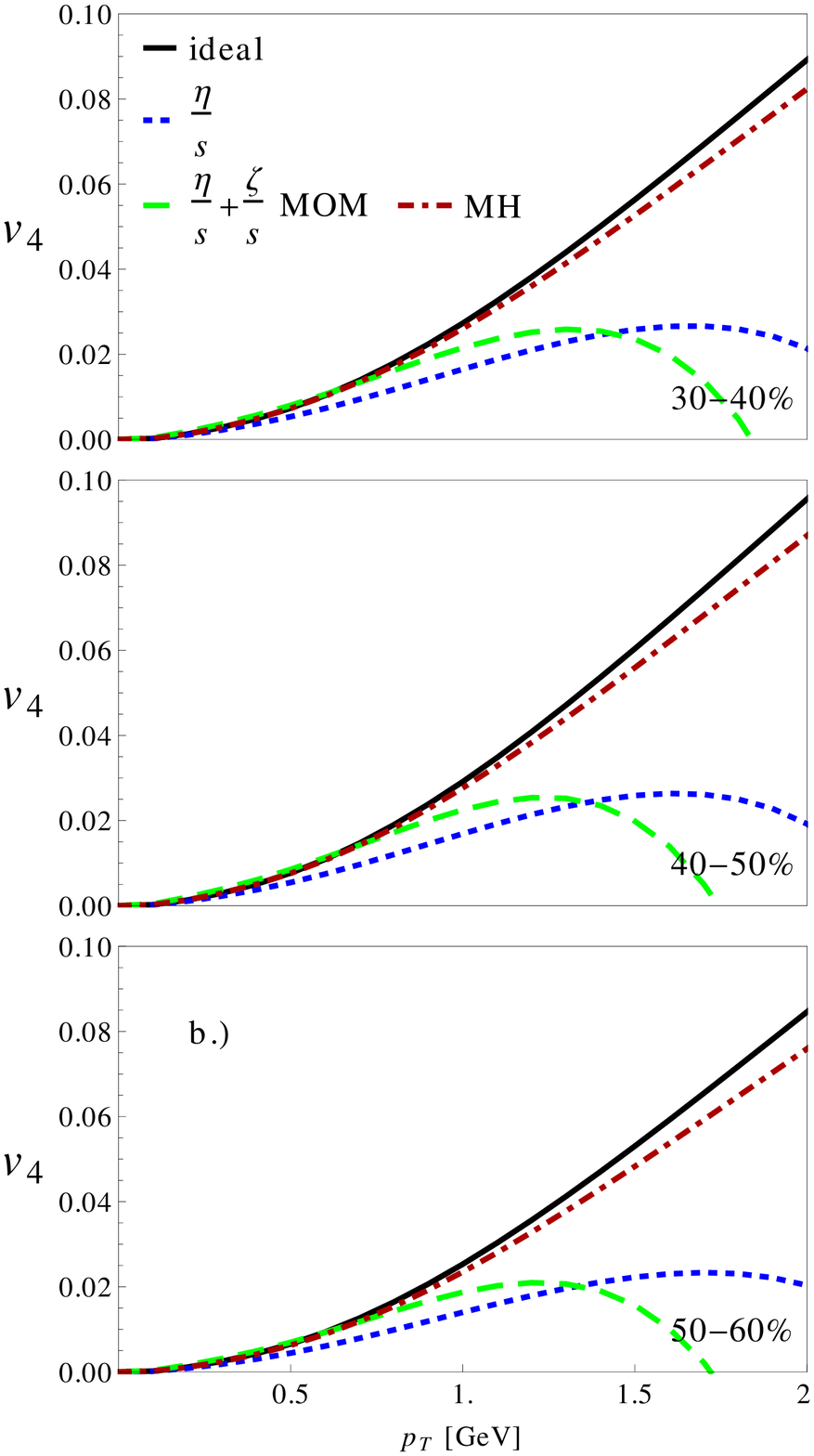} \\ 
\end{tabular}%
\caption{(Color online) Comparison of $v_4$ of direct pions across the centrality classes $0-10\%$, $10-20\%$, $20-30\%$ (in column a.), $30-40\%$, $40-50\%$, and $50-60\%$ (in column b.) for RHIC. The solid black line denotes the ideal hydro result, the short-dashed blue line was computed taking into account only shear viscosity, the long-dashed green curve was computed using shear+bulk with the moments method expression for the freeze-out while the dark red dotted-dashed curve corresponds to shear+bulk with the MH formula.}
\label{tab:v4}
\end{figure}

\begin{figure}[ht]
\centering
\begin{tabular}{cc}
\includegraphics[width=0.5\textwidth]{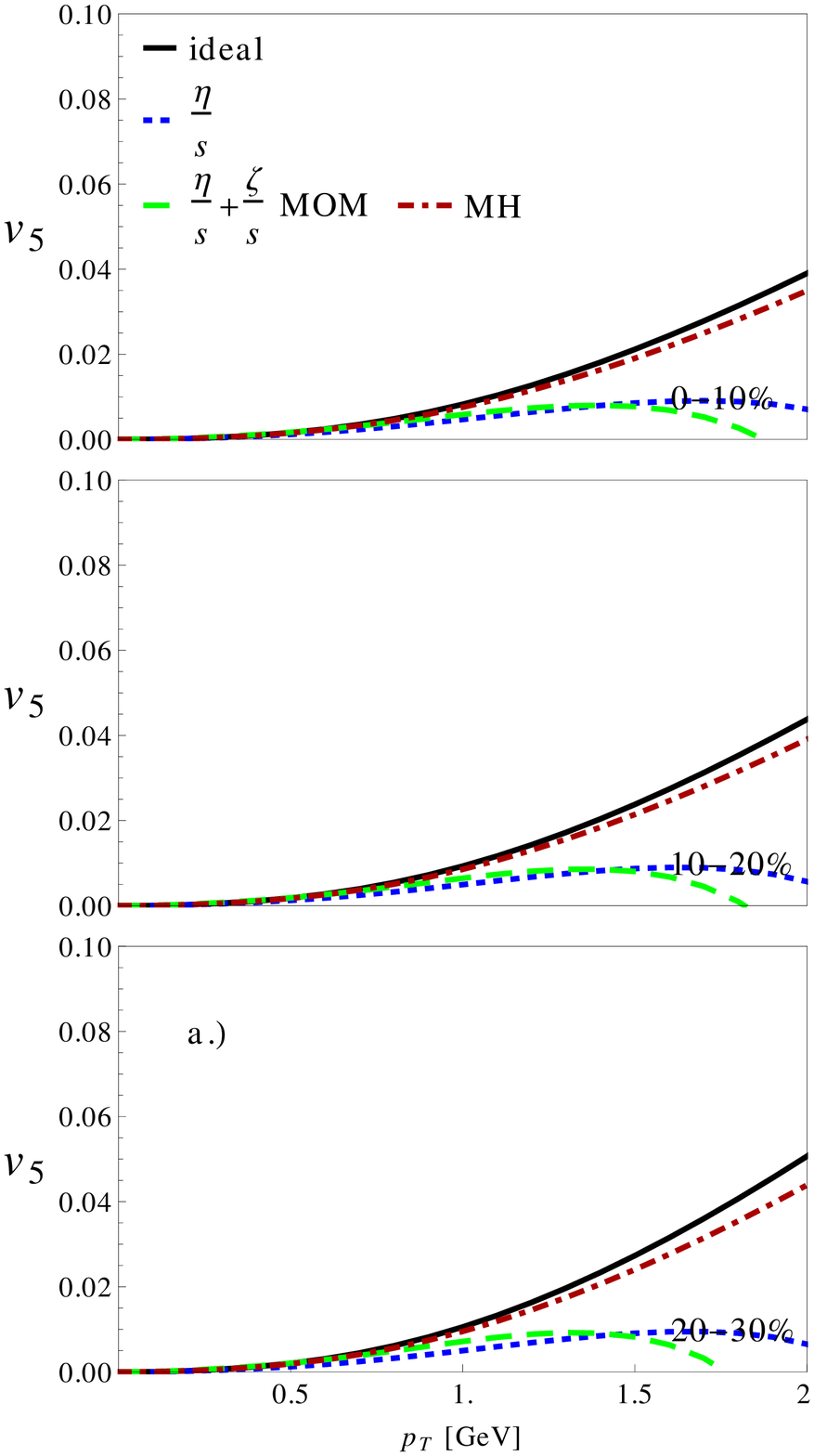} & %
\includegraphics[width=0.5\textwidth]{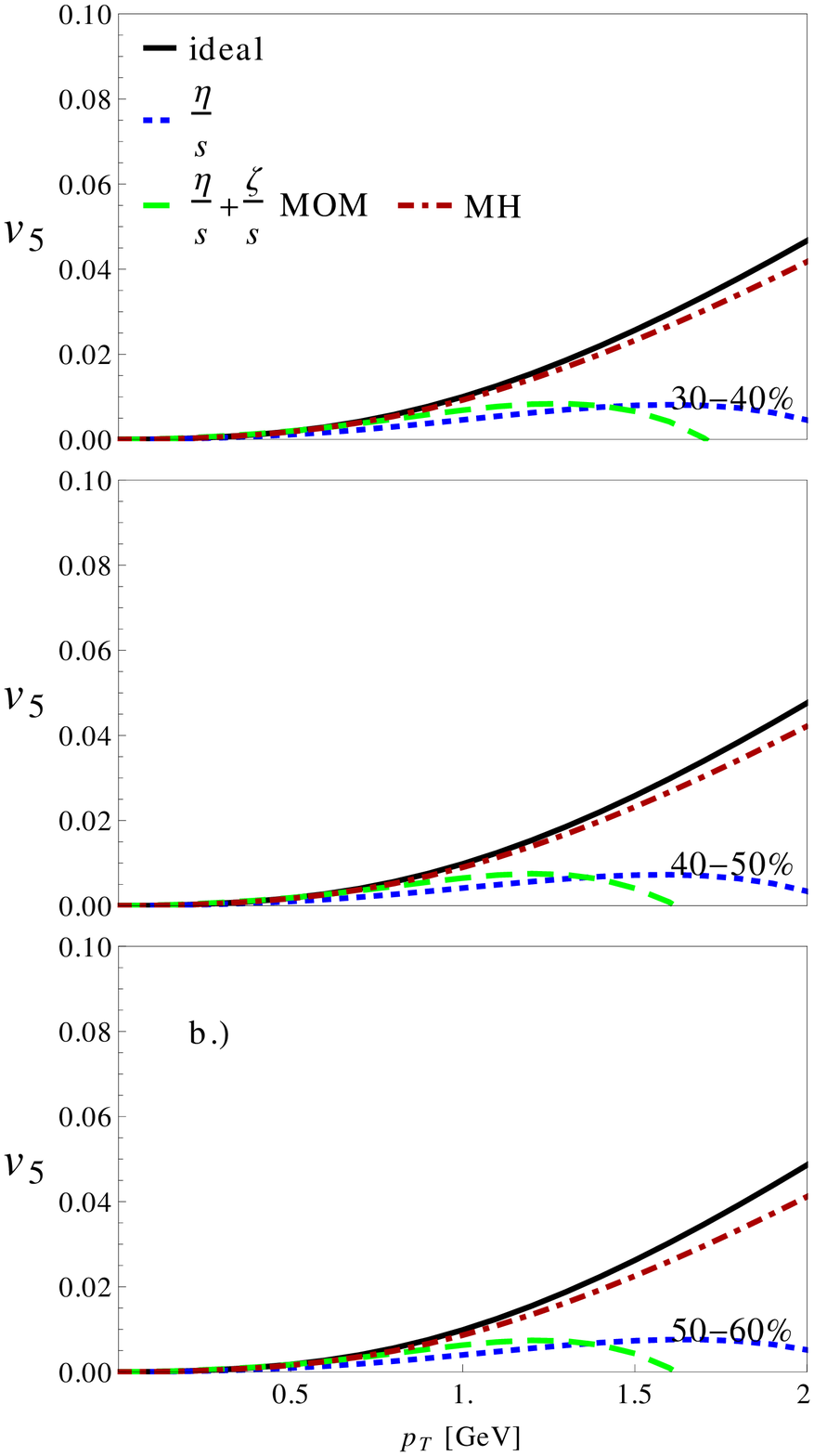} \\ 
\end{tabular}%
\caption{(Color online) Comparison of $v_5$ of direct pions across the centrality classes $0-10\%$, $10-20\%$, $20-30\%$ (in column a.), $30-40\%$, $40-50\%$, and $50-60\%$ (in column b.) for RHIC. The solid black line denotes the ideal hydro result, the short-dashed blue line was computed taking into account only shear viscosity, the long-dashed green curve was computed using shear+bulk with the moments method expression for the freeze-out while the dark red dotted-dashed curve corresponds to shear+bulk with the MH formula.}
\label{tab:v5}
\end{figure}

In this section we consider the $p_T$ dependent $v_n$'s across all centrality classes and look for the effects of bulk and shear viscosity.  In Figs.\ \ref{tab:v2}-\ref{tab:v5} we show $v_2(p_T)$-$v_5(p_T)$, respectively.  We do not include the effect of $\eta/s+10\,\zeta/s$ because for this large $\zeta/s$ value we are only able to calculate the spectrum reliably up to $p_T < 1$ GeV for MOM (in the case of MH one can integrate up to $p_T< 1.5$ GeV). One can see that the viscous corrections considerably change the $v_n(p_T)$'s, especially for large $p_T$.  For all the $v_n(p_T)$'s and all the centralities we see a pattern: ideal hydrodynamics (solid black line) gives the largest $v_n(p_T)$, followed by the viscous case with shear and bulk computed using the MH formula (dark red dotted-dashed curve) and then the shear+bulk computed using the MOM expression (long-dashed green line), for which the $v_n(p_T)$'s are a bit larger than the case including only shear viscosity (short-dashed blue line) for $p_T < 1.5$ GeV. These results show that the bulk viscosity-driven suppression of shear effects also occurs for the differential flow harmonics.

In the previous section we showed that both MH and MOM give very similar integrated $v_n$'s. The same cannot be said about the $p_T$ differential flow harmonics. Using the MOM correction, we found that the case including $\eta/s+\zeta/s$ effects uniformly decreases the $v_n(p_T)$'s and the effect is strongest for the most peripheral collisions.  Additionally, higher order $v_n$'s are more strongly affected by the combined effect of shear and bulk viscosities. We also note that for $v_2(p_T)$ the MH curves with shear and bulk nearly match the ideal curves for all centrality classes. The same does not occur for higher order flow coefficients. In fact, as discussed in \cite{Noronha-Hostler:2013gga}, the MH correction to the particle distribution diverges quickly for large $p_T$. The curves computed with the MOM method start to decrease (faster than those for pure shear viscosity effects) for $p_T >1.5$ GeV.

\subsection{Equal Shear and Bulk viscosities}\label{eqn:equal}

Up until this point we have always assumed that the bulk viscosity is significantly smaller than the shear viscosity.  However, due to our very limited knowledge about the magnitude of $\zeta/s$ in the QGP there is no a priori reason why that must be the case. While there are no limitations from the point of view of the hydrodynamic code to use larger $\zeta/s$'s (aside from possible cavitation effects for sufficiently large $\zeta/s$), unfortunately, due to limitations with the $\delta f$ corrections for bulk viscosity we cannot include a bulk viscosity that is as large as the generally accepted shear viscosity $\sim 1/4\pi$. However, in order to understand what happens when both the bulk and shear viscosities have equal magnitude, we can consider a very small shear viscosity that is of the same order of magnitude as our bulk viscosity.  In this section we consider the temperature independent situation where $\zeta/s=\eta/s=0.007$ and compare to the case where only shear viscosity $\eta/s=0.007$ is included.  Because the bulk viscosity generally increases the $v_n$'s (both integrated and $p_T$ dependent flow harmonics) and the shear viscosity generally decreases them, it is possible that when they are both of the same order of magnitude that they will reproduce the ideal results (an indication of that possibility was already found in the previous section when comparing the shear+bulk MH results for $v_2(p_T)$ with its ideal value).  Here we only look at the $20-30\%$ centrality class.

\begin{figure}[ht]
\includegraphics[width=0.5\textwidth]{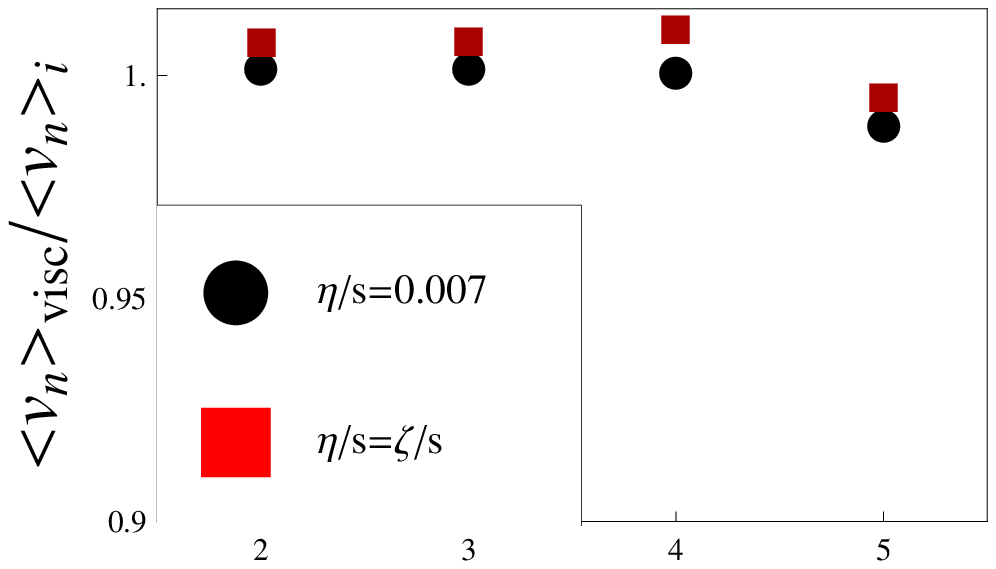} 
\caption{(Color online) Ratio between the integrated $v_n$'s of viscous and ideal hydrodynamics of direct pions in the $20-30\%$ centrality class at RHIC computed using the Moments Method (MOM) for the pure shear case $\eta/s=0.007$ (black dots) and the bulk + shear calculation where $\zeta/s=\eta/s=0.007$ (red squares). }
\label{fig:vnequal}
\end{figure}

In Fig.\ \ref{fig:vnequal} the integrated $v_n$'s for direct pions are shown for RHIC's $20-30\%$ most central collisions. While such a small shear viscosity of $\eta/s=0.007$ has an extremely small effect on the integrated $v_n$'s, one can still see that it does decrease the $v_n$'s and the higher order $n$'s are most strongly affected.  However, when bulk viscosity is included we see that the integrated $v_n$'s return to almost precisely the result for ideal hydrodynamics (with the exception of $v_5$, which remains slightly below still).  This indicates that it may be possible for bulk viscosity to compensate for the effects of shear viscosity in the integrated $v_n$'s when they are both of the same order of magnitude.

\begin{figure}[ht]
\centering
\begin{tabular}{cc}
\includegraphics[width=0.5\textwidth]{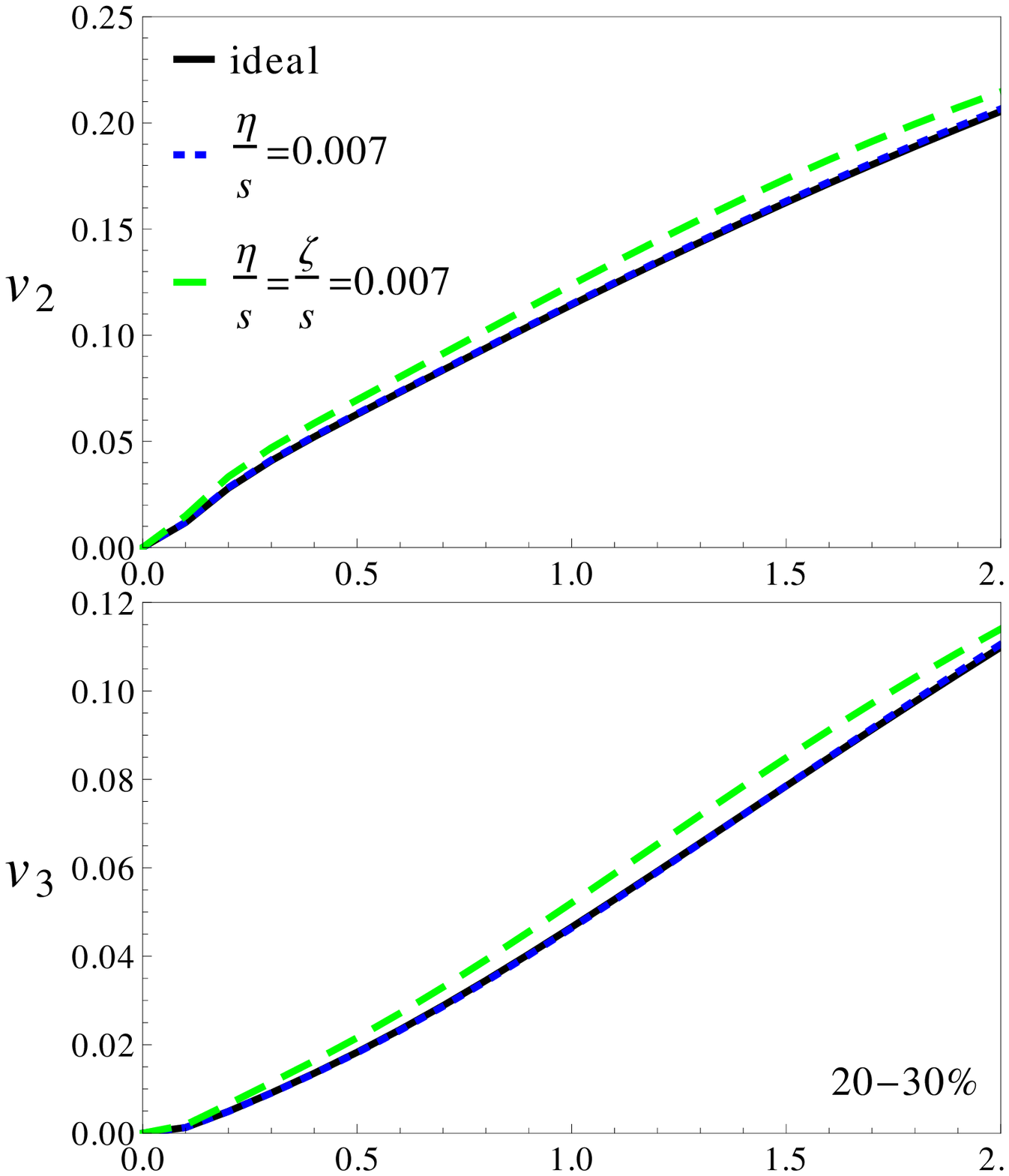} & %
\includegraphics[width=0.5\textwidth]{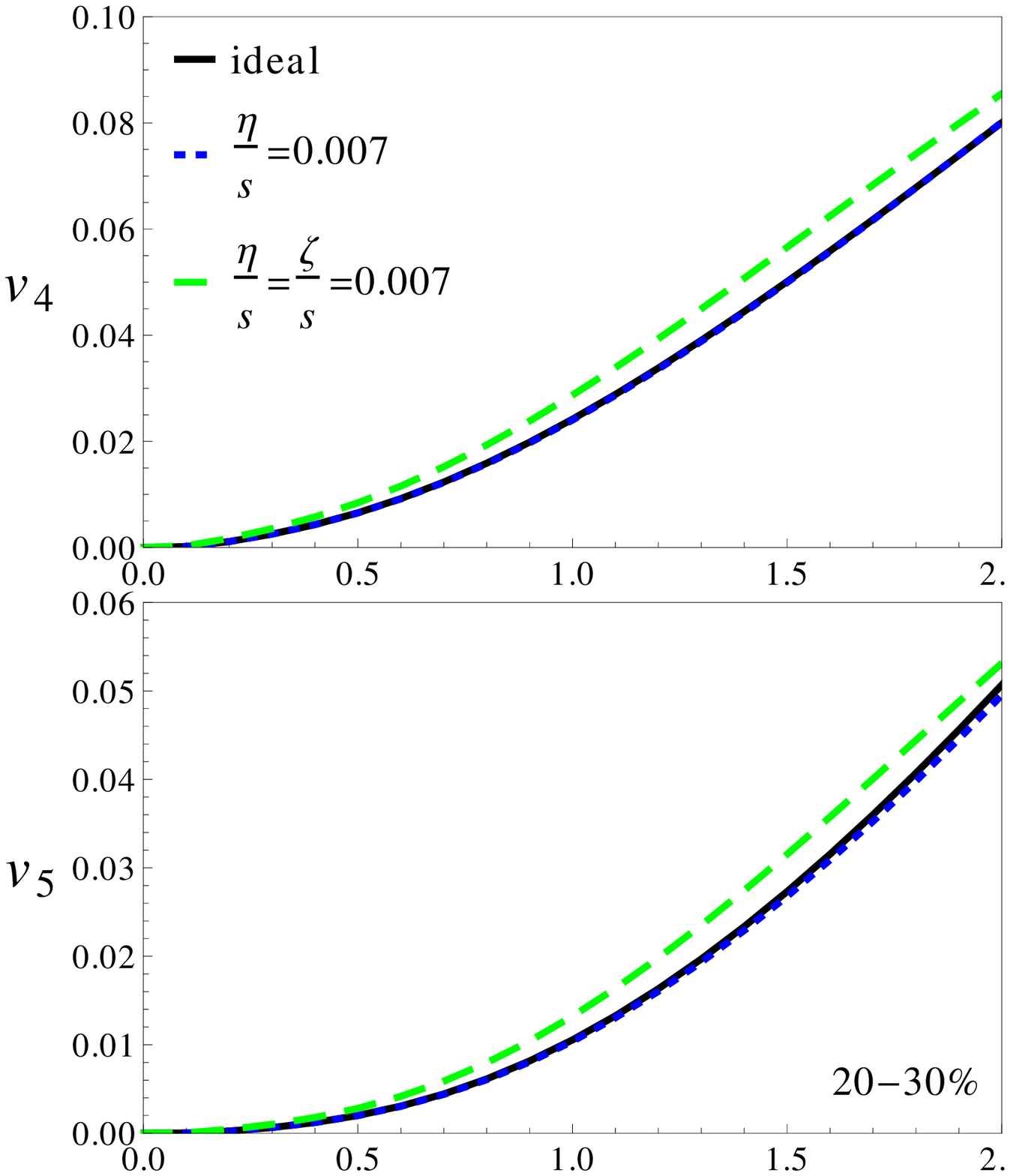} \\ 
\end{tabular}%
\caption{(Color online) Comparison between the $v_n(p_T)$'s of direct pions for the centrality class $20-30\%$ at RHIC when bulk and shear viscosities have equal magnitude. The solid black curves denote the ideal hydrodynamics results, the short-dashed blue curve shows the result in the case where there is only shear viscosity $\eta/s=0.007$ while the long-dashed green curve corresponds to the case where $\zeta/s=\eta/s=0.007$ computed using the MOM approach.}
\label{fig:vnequalpt}
\end{figure}
In Fig.\ \ref{fig:vnequalpt} we observe the $p_T$ dependent flow harmonics of direct pions in the centrality class $20-30\%$ for the small $\eta/s$ case.  One can see that there is no visible difference between the ideal case and that of a small shear viscosity of $\eta/s=0.007$. However, when the bulk and shear viscosities are identical the $p_T$ dependent $v_n$'s increase, which indicates that the effect of bulk viscosity dominates the $p_T$ dependent $v_n$'s when bulk and shear are of the same order of magnitude.  This was hinted already in Fig.\ \ref{tab:vintmom} when we used $10\zeta/s$. However, in that case we were limited to a small $p_T$ range over which we could integrate our spectrum due to the problems of the $\delta f$.  Here we avoid that problem due to the small value of $\zeta/s$ and $\eta/s$.

\section{Conclusions}\label{sec:conclu}

In this paper we used v-USPhydro (a 2+1 Lagrangian hydrodynamical model) to study the effects of both bulk and shear viscosities on the hydrodynamical evolution of the QGP and the resulting anisotropic collective flow harmonics at RHIC. We found that even though in our equations of motion the shear stress tensor $\pi^{\mu\nu}$ and the bulk scalar $\Pi$ do not couple directly, their indirect coupling via the flow velocity is still strong enough for them to influence each other in a nonlinear fashion. We found that the inclusion of even a small bulk viscosity decreases the well-known shear viscosity-induced suppression of both the integrated and differential flow harmonics bringing them closer to their values in ideal hydrodynamical calculations. This is a new effect brought in by bulk viscosity in event by event hydrodynamic simulations. 

Furthermore, we found that when the bulk and shear viscosities are roughly the same order of magnitude the bulk viscosity negates a significant portion of the contribution from shear viscosity for the integrated flow harmonics. In fact, for the differential flow harmonics the bulk viscosity dominates and actually increases the flow harmonics, which is the opposite effect when only the shear viscosity is considered. Even if one considers only a small $\zeta/s$ we find that the bulk viscosity tempers the suppression typically demonstrated from the shear viscosity. When one looks only at the relativistic hydrodynamical expansion one can clearly see that the presence of bulk viscosity suppresses the components of the shear stress tensor.  Additionally, we find that this effect plays the largest role the longer the system is expanding. Thus, for the case of lower freeze-out temperatures and also for higher collision energies where one expects to see more time spent within the hydrodynamical expansion, one would expect that the effects of bulk viscosity are more relevant. 

The effects of viscosity are most relevant for peripheral collisions and higher order flow harmonics for both the integrated and differential $v_n$'s.  Furthermore, we find that for the integrated $v_n$'s when we include a large $\zeta/s$ that $v_2\approx v_3$ for the most central collisions but that $v_3$ is more significantly suppressed for more peripheral collisions. When $\eta/s=\zeta/s$ (but both are small) $v_4$ basically returns to its value in ideal hydrodynamics while the same is not true for $v_5$. This could indicate that the higher order flow harmonics may be vital in helping us estimate $\eta/s$ and $\zeta/s$ within relativistic heavy ion collisions because they do not allow for a complete compensation of shear viscosity effects due to bulk viscosity. 

Our calculations need to be improved in a number of ways. First, in this paper we did not look into the flow harmonics of hadronic species other than pions and no particle decays or hadronic afterburner effects have been included. Clearly, this must be improved to allow for a comparison to data and adequately evaluate the role played by bulk viscosity in the flow harmonics of the QGP formed in heavy ions collisions. Also, different set of initial conditions (we have only used MC Glauber in this paper) and different collisions systems and energies should be investigated. Furthermore, the considerable difference found in the differential anisotropic flow coefficients computed using two different $\delta f$'s formulas is an issue that should serve as a motivation for finding a better behaved expression for the species dependent viscous corrections at freeze-out including both bulk and shear effects. Moreover, as discussed in \cite{Dumitru:2007qr}, constraints on the entropy production generate a correlation between the values of transport coefficients and the initial time for hydrodynamics. If bulk viscosity effects compensate the effects from shear, $\eta/s$ could be larger than used here and, consequently, $\tau_0$ may actually be larger than usually considered in hydrodynamic simulations.

As mentioned above, the bulk viscosity-driven suppression of the shear stress tensor found here occurs in a very indirect way mediated by the modification of the flow velocity due to bulk viscosity. Indeed, in our equations of motion we do not include the known terms \cite{Denicol:2012cn} which display a direct coupling between $\Pi$ and $\pi^{\mu\nu}$. It would be interesting to see if the effect discussed here remains when the more general equations of motion of \cite{Denicol:2012cn} and transport coefficients \cite{Denicol:2014vaa} are used in the hydrodynamical evolution. We hope to address this question in the near future \cite{comment}.

We remark that the actual magnitude (and temperature dependence) of $\zeta/s$ in heavy ion collisions is largely unknown and it is conceivable that depending on the value of $\zeta/s$ in the QGP, the bulk viscosity-driven suppression of shear viscosity effects on the flow harmonics found here may require a re-evaluation of the previous estimates of $\eta/s$ extracted from comparisons of hydrodynamic calculations (which did not include bulk viscosity effects) to heavy ion data.

\section*{Acknowledgements}
We thank G.~S.~Denicol for helping us with the implementation of shear viscosity effects in the v-USPhydro code. We thank G.~Torrieri for insightful discussions about the effects of bulk viscosity in heavy ion collisions and A.~Dumitru for comments about the thermalization time and the values of transport coefficients. This work was supported by Funda\c c\~ao de Amparo \`a Pesquisa do Estado de S\~ao Paulo (FAPESP) and Conselho Nacional de Desenvolvimento Cient\'ifico e
Tecnol\'ogico (CNPq). This work was also financially supported by the Helmholtz International Center for FAIR within the
(Landesoffensive zur Entwicklung Wissenschaftlich-konomischer Exzellenz) launched by the State of Hesse.

\appendix

\section{Tests of the numerical code}

Because of the complexity of viscous relativistic hydrodynamical equations, it is vital to test the accuracy of our numerical code.  Fortunately, there are now well-known numerical and semi-analytical solutions for this purpose.

\subsection{TECHQM}

\begin{figure}[ht]
\centering
\begin{tabular}{cc}
\includegraphics[width=0.5\textwidth]{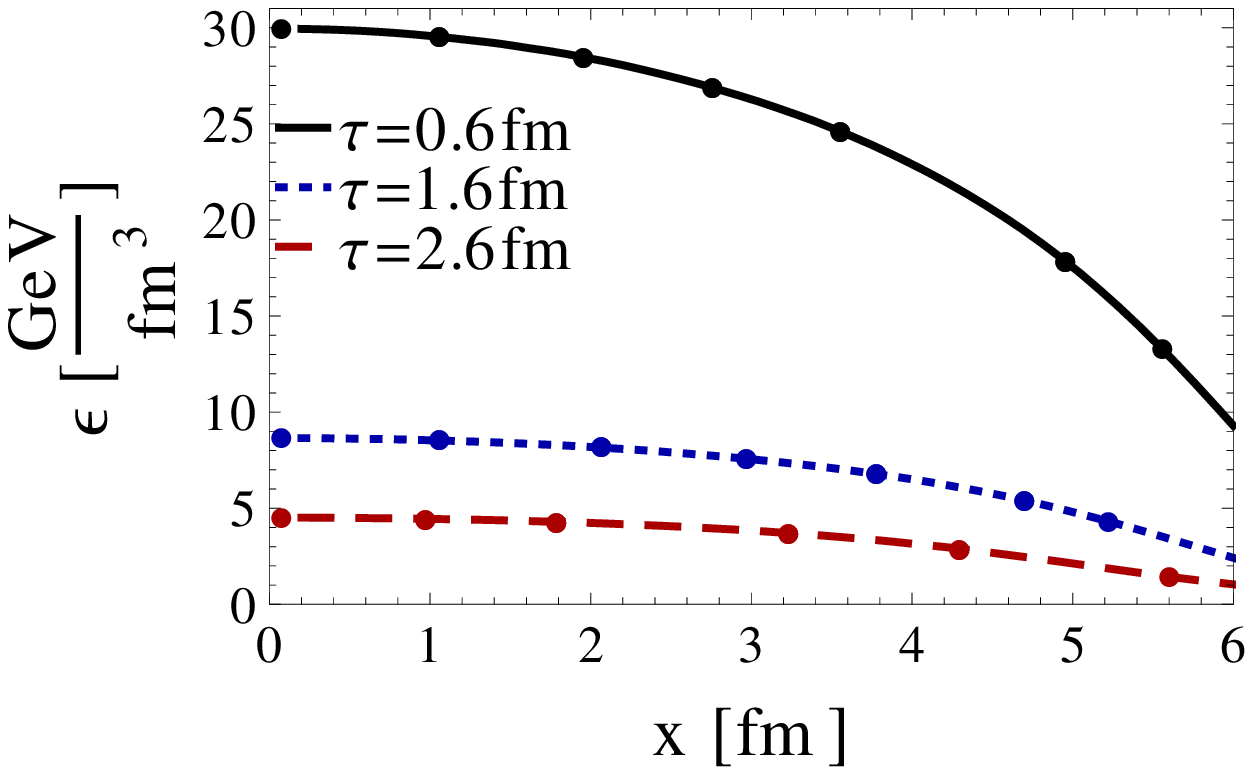} & %
\includegraphics[width=0.5\textwidth]{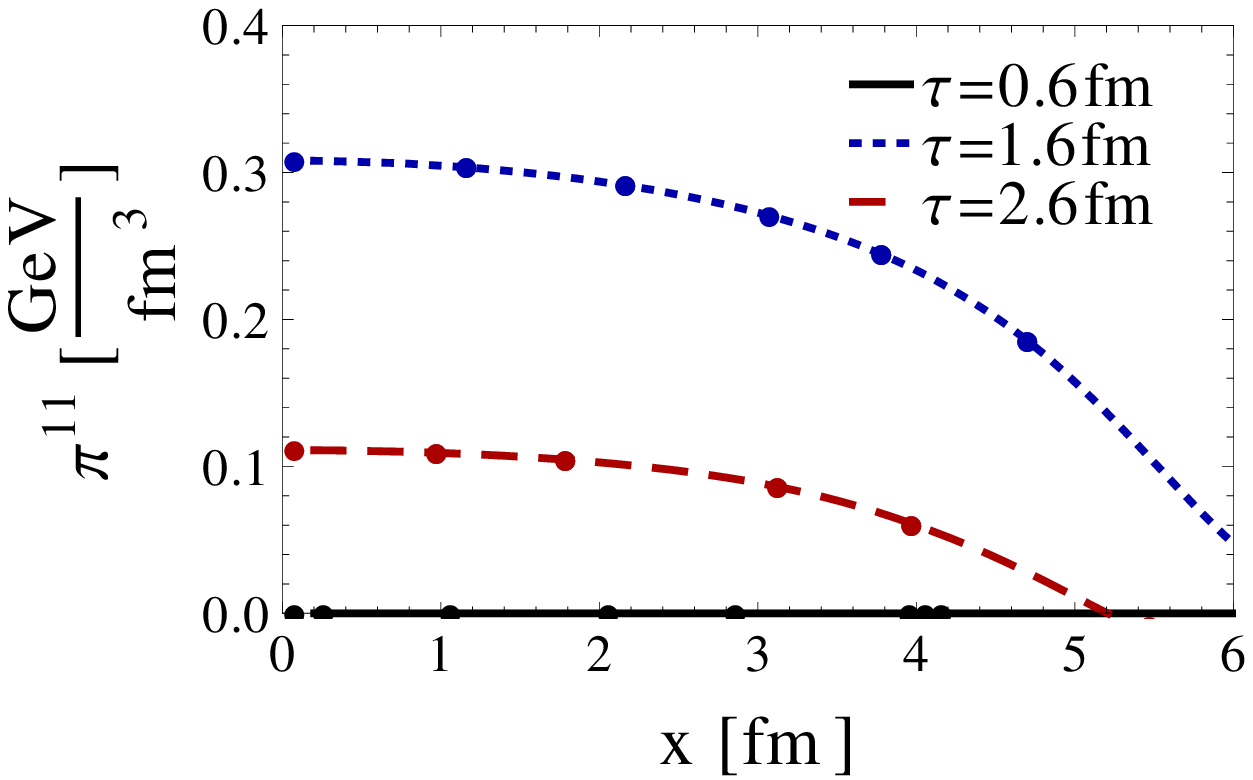} \\ 
\newline
\includegraphics[width=0.5\textwidth]{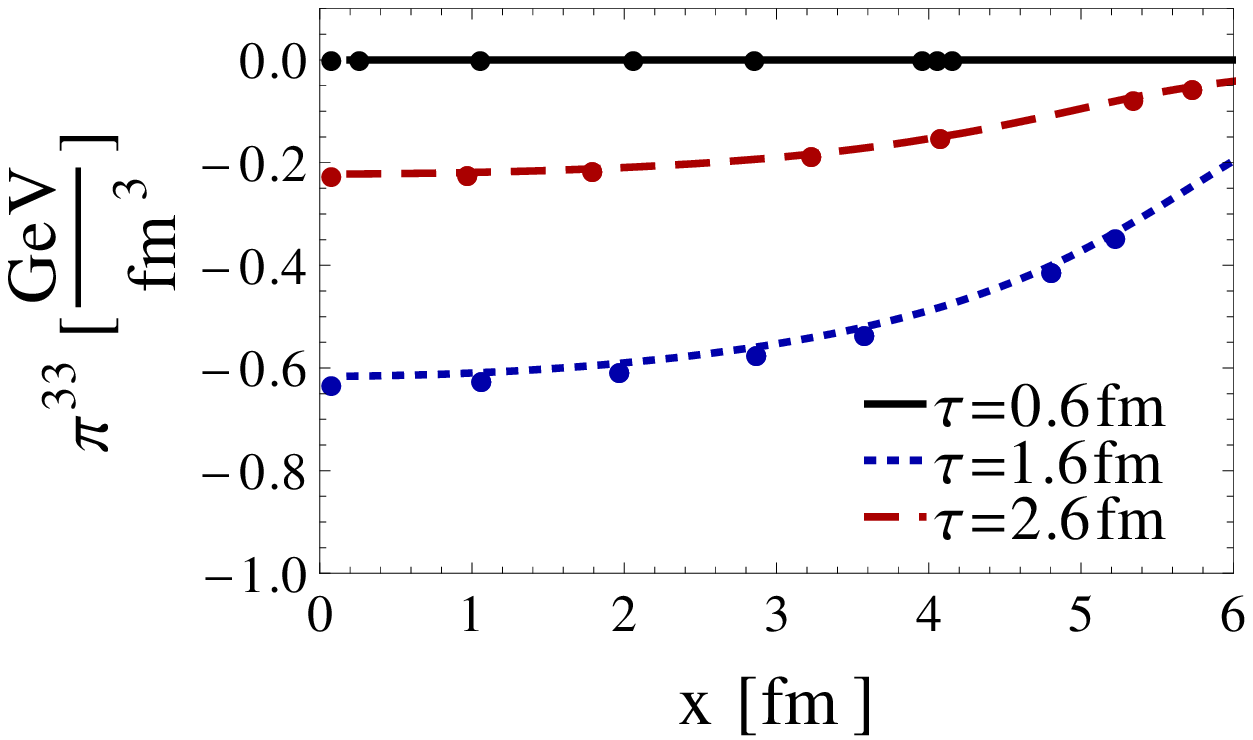} & %
\includegraphics[width=0.5\textwidth]{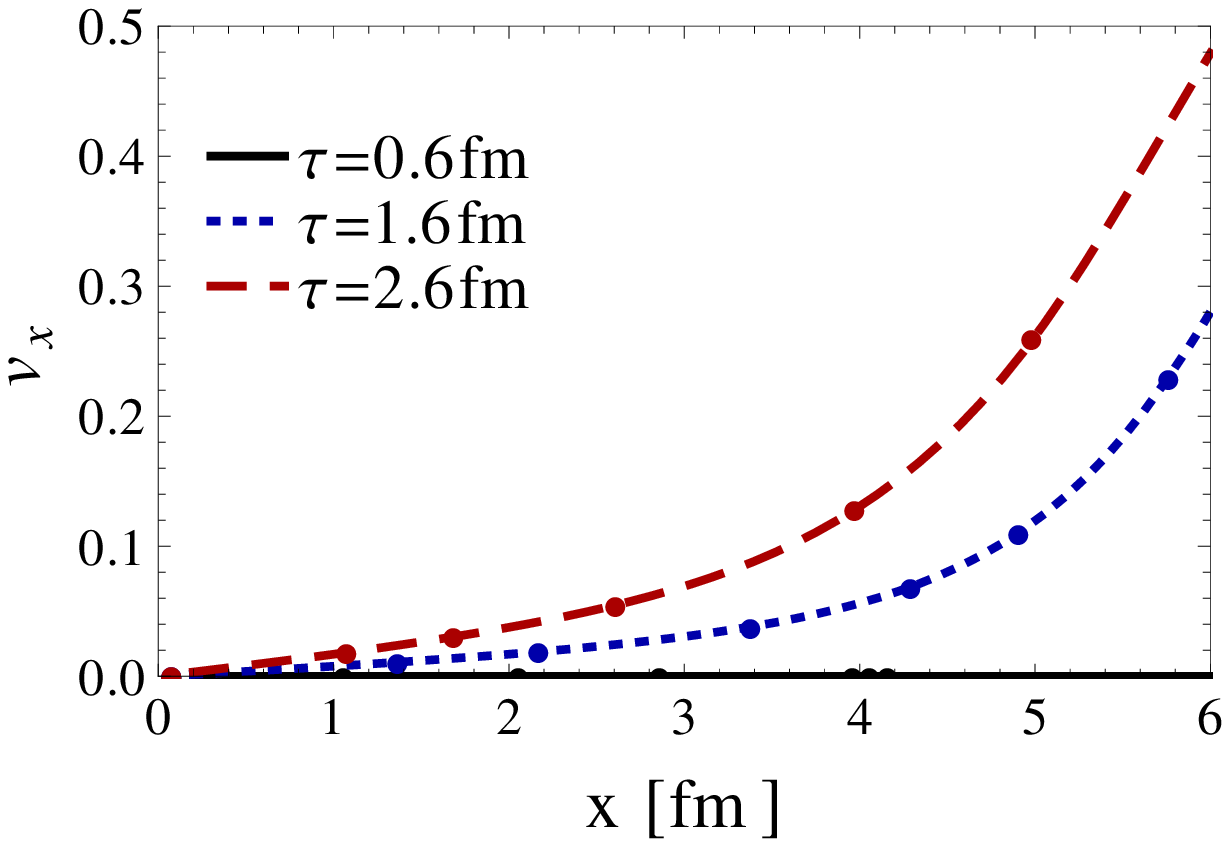} \\ 
\end{tabular}%
\caption{(Color online) Comparison between the TECHQM results (lines) and our results (dots) for the energy density $\varepsilon$, the shear stress components $\pi^{11}$ and $\pi^{33}$, as well as the x-component of the flow velocity. The comparison is made at the times $\tau=0.6$ (solid black lines), $1.6$ (blue short dashed lines), and $2.6$ fm (red long dashed lines).}
\label{tab:techqm}
\end{figure}

One aspect of TECHQM bulk evolution working group \cite{techqm} was to ensure the overall accuracy of relativistic hydrodynamical codes and solutions for both ideal and viscous hydro evolution (with only shear viscosity) have become available. We use an ideal equation of state and the $b=0$ fm central Au-Au optical Glauber initial condition at RHIC $\sqrt{s}=200$ A GeV.  We take the starting time for hydrodynamics to be $\tau_0=0.6$ fm, $\eta/s=0.08$, $\tau_{\pi}=3\eta/(sT)$, and the freeze-out temperature is $T_0=130$ MeV.  In Fig.\ \ref{tab:techqm} we compare our results (dots) to the TECHQM results (lines) \cite{techqm} over time $\tau=0.6$ (solid black lines), $1.6$ (blue short dashed lines), and $2.6$ fm (red long dashed lines).  One can clearly see that the results match well for multiple time steps.  In our code we used the SPH smoothing parameter is $h=0.2$ fm, the total number of SPH particles is $N_{SPH}= 25432$, and the time step $d\tau=0.02$ fm.

\subsection{$SO(3)\otimes SO(1,1)\otimes Z_2$ Test of Conformal Israel-Stewart Dynamics}\label{sheartest}

Using symmetry arguments, Gubser \cite{Gubser:2010ze} found analytical solutions of ideal and Navier-Stokes conformal hydrodynamics that are invariant under $SO(3)\otimes SO(1,1)\otimes Z_2$ (a subgroup of $SO(2,4)$). The symmetries imply that the solution is radially symmetric in the transverse plane and boost invariant with the flow
\begin{eqnarray}
u_{\tau}&=&-\cosh \left[\tanh^{-1}\left(\frac{2q^2\tau r}{1+q^2\tau^2+q^2 r^2}\right)\right]\nonumber\\
u_{r}&=&\sinh \left[\tanh^{-1}\left(\frac{2q^2\tau r}{1+q^2\tau^2+q^2 r^2}\right)\right]\nonumber\\
u_{\phi}&=&u_{\eta}=0
\end{eqnarray}
where $q$ is a free parameter with dimensions of energy. This approach has been used in \cite{Marrochio:2013wla} to find the first analytical and semi-analytical solutions of conformal Israel-Stewart hydrodynamics which includes nontrivial dynamics in the transverse plane. These solutions are described in detail in \cite{Marrochio:2013wla} and they provide a very stringent test of the accuracy of viscous hydrodynamic codes. Since the equations of motion involving shear viscosity used here have the same structure as those in \cite{Marrochio:2013wla}, we can directly test the accuracy of v-USPhydro in this case. We also note that novel analytical solutions of conformal Israel-Stewart hydrodynamics with full 3+1 dynamics can be found in \cite{Hatta:2014gqa,Hatta:2014gga}. 

\begin{figure}[ht]
\centering
\begin{tabular}{cc}
\includegraphics[width=0.5\textwidth]{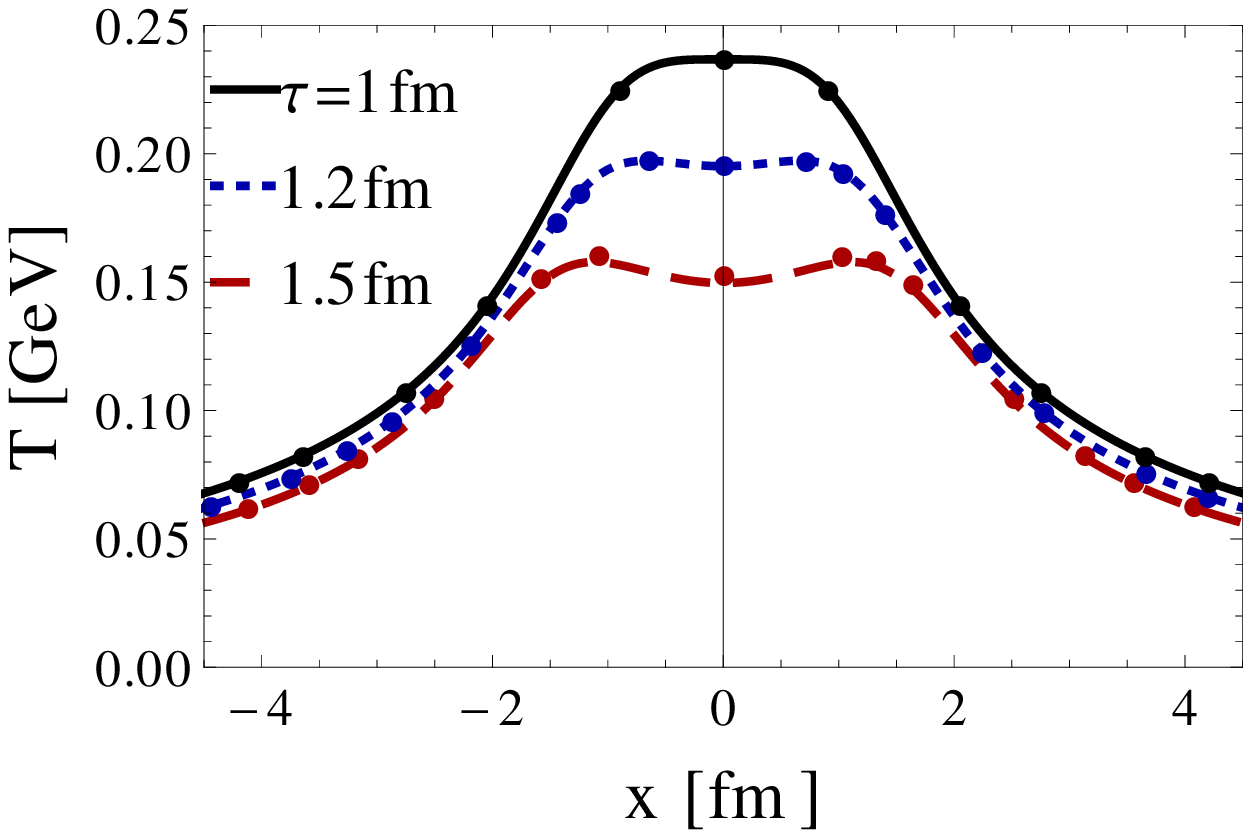} & %
\includegraphics[width=0.5\textwidth]{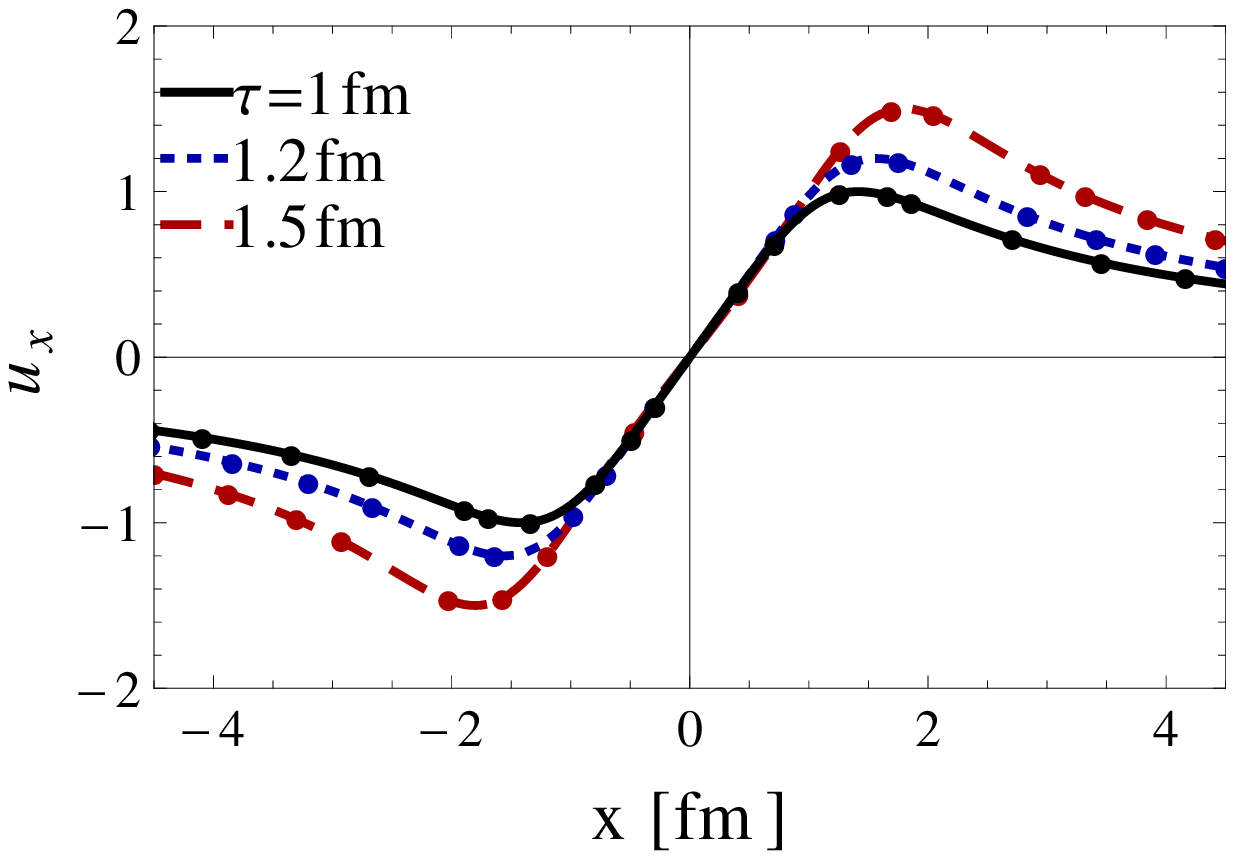} \\ 
\newline
\includegraphics[width=0.5\textwidth]{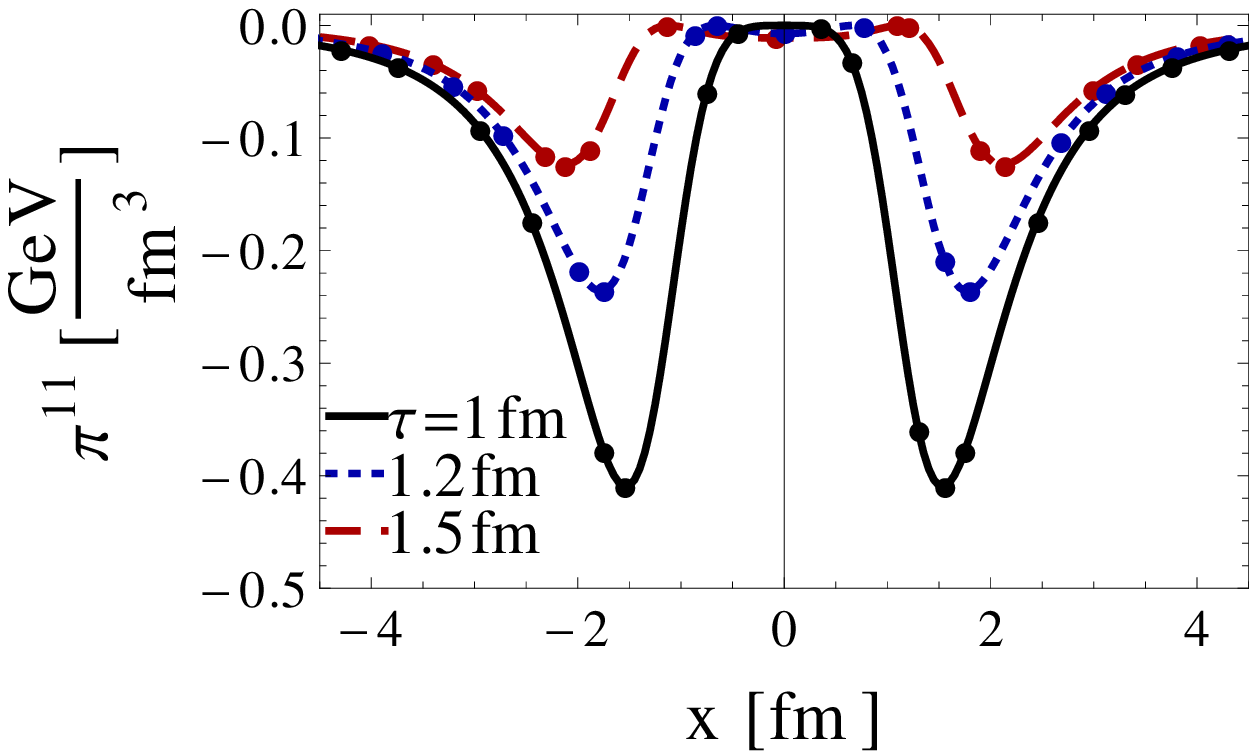} & %
\includegraphics[width=0.5\textwidth]{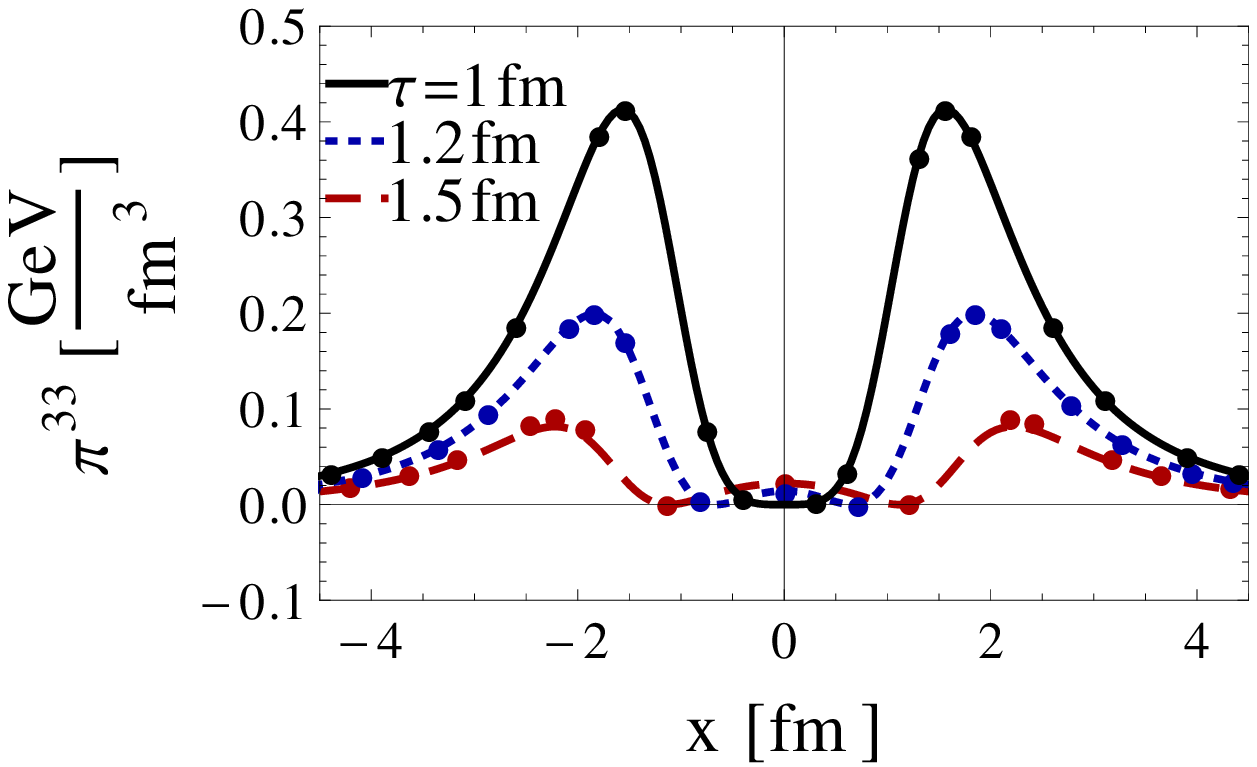} \\ 
\end{tabular}%
\caption{(Color online) Comparison between the semi-analytical solutions of \cite{Marrochio:2013wla} ( lines) and v-USPhydro (dots). Here we compare for the time steps $\tau=1.0$ fm (first time step, which sets the initial condition for the fields) (solid black lines), $1.2$ fm (blue short dashed lines), and $1.5$ fm (red long dashed lines).}
\label{tab:gubser}
\end{figure}

Here we compare the results from v-USPhydro to the semi-analytical solution \cite{Marrochio:2013wla} in the case where $\eta/s=0.2$, $\tau_\pi=5(\eta/s)/T$, and $q=1$ fm$^{-1}$. The comparison involving the temperature, the flow, and a few components of the shear stress tensor can be found in Fig.\ \ref{tab:gubser}.  For our comparison we used $h=0.1$, $d\tau=0.001$ fm, $\tau_0=1$ fm, and  the total number of SPH particles is $N_{SPH}= 40401$.  We see that v-USPhydro is able to match the semi-analytical solution pretty well at early times (the agreement remains at later times). 

\section{Details of the Cooper-Frye freeze-out in the SPH formalism}\label{detailsCF}

Our distribution function for a single particle species that includes effects from both shear and bulk viscosity is
\begin{equation}
f_{p}=f_{0p}\left[1+\left(1-af_{0p}\right)\left(\delta f^{Bulk}_{p}+\delta f^{Shear}_{p}\right)\right]
\end{equation}
where $p^\mu$ is the particle on-shell momentum. The ideal distribution, $f_{0p}$, is defined as
\begin{equation}
f_{0p}=\frac{1}{e^{(p^{\mu}u_{\mu})/T_0}+a}
\end{equation}
where $a=1$ for fermions, $-1$ for bosons, and $0$ for classical Boltzmann statistics. For the ideal component we have
\begin{equation}
f_{0p}=\sum_{n=0}^{\infty}\left(-a\right)^n e^{-(n+1)\frac{p^{\mu}u_{\mu}}{T_0}}
\end{equation}
whereas
\begin{equation}
f_{0p}\left(1-af_{0p}\right)=\sum_{n=0}^{\infty}\left( n+1\right)\left(-a_i\right)^n e^{-(n+1)\frac{p^{\mu}u_{\mu}}{T_0}}
\end{equation}
and
\begin{equation}\label{eqn:cordis}
f_{p}=\sum_{n=0}^{\infty}\left(-a\right)^n e^{-(n+1)\frac{p^{\mu}u_{\mu}}{T_0}}\left(1+  \left( n+1\right) \left[\delta f^{Bulk}_{p}+\delta f^{Shear}_{p}\right]\right)\,.
\end{equation}

In Cartesian coordinates the scalar product of the particle momentum and the hypersurface normal vector (see Appendix \ref{normalvector}), $n^\mu$, is
\begin{equation}
p^{\mu}\cdot n_{\mu}=E n_t+p^x n_x+p^y n_y +p^z n_z
\end{equation}
switching to hyperbolic coordinates
\begin{eqnarray}
p^{\mu}\cdot u_{\mu}&=&  m_{\perp} u_{\tau} \text{cosh}\left(\eta-y\right) - \vec{p}_{\perp}\cdot \vec{u}_{\perp}  \\
p^{\mu}\cdot n_{\mu}&=&  m_{\perp} n_{\tau} \text{cosh}\left(\eta-y\right) - \vec{p}_{\perp}\cdot \vec{n}_{\perp} \\
u^{\mu} \cdot n_{\mu}&=& u_{\tau}n_{\tau}+u^{x}n_x+u^y n_y\,.
\end{eqnarray}
In the SPH formalism the integral over the isothermal hypersurface is written in terms
of a sum of SPH particles as 
\begin{equation}\label{spectraSPH}
\frac{dN}{dyd^2p_T} = \frac{g}{(2\pi)^3}\,\sum_{\alpha=1}^{N_{SPH}}%
\int_{-\infty}^{\infty}d\eta_\alpha\frac{(p\cdot n)_\alpha}{ (n\cdot u)_\alpha}\frac{\nu_\alpha}{\sigma_\alpha}
\,f(T_{FO}, (p\cdot u)_\alpha,\Pi_\alpha,\pi^{\mu\nu}_\alpha)
\end{equation}
where $N_{SPH}$ is the total
number of SPH particles, $(n_\mu)_\alpha$ is the normal vector of the
isothermal hypersurface reconstructed using the $\alpha$-th SPH particle, $%
(u_\mu)_\alpha$ is the 4-velocity of the SPH particle,  $\Pi_\alpha$ is
the bulk viscosity of the SPH particle, there is an integral over the space-time rapidity of each SPH particle $\eta_\alpha$, and $\pi^{\mu\nu}_\alpha$ is
the shear stress tensor of the SPH particle. 

Then, the contribution from a SPH particle to the ideal distribution function is
\begin{equation}
f^{(\alpha)}_{0p}=e^{ \vec{p}_{\perp}\cdot \vec{u}^{(\alpha)}_{\perp}/T_0}e^{- m_{\perp} u^{(\alpha)}_{\tau}/T_0 \text{cosh}\left(\eta_\alpha-y\right)}
\end{equation}
Substituting this into Eq.\ (\ref{eqn:cordis}), we find
\begin{equation}\label{eqn:cordiswithns}
f^{(\alpha)}_{p}=\sum_{n=0}^{\infty}\left(-a_\alpha\right)^n \lambda_\alpha^{n+1} e^{-(n+1)m_{\perp} u^{(\alpha)}_{\tau}/T_0 \text{cosh}\left(\eta_\alpha-y\right)}\left(1+  \left( n+1\right) \left[\delta f^{(\alpha)Bulk}_{p}+\delta f^{(\alpha)Shear}_{p}\right]\right)
\end{equation}
where $\lambda_\alpha=e^{ \vec{p}_{\perp}\cdot \vec{u}^{(\alpha)}_{\perp}/T_0}$.

The integral over the isothermal hypersurface becomes
\begin{eqnarray}
\frac{dN}{dyd^2p_T}& =& \frac{g}{(2\pi)^3}\,\sum_{\alpha=1}^{N_{SPH}}\frac{1}{ (n\cdot u)_\alpha} \frac{\nu_\alpha}{\sigma_\alpha} \left\{  m_{\perp} n_{\tau}^{(\alpha)} 
\int_{-\infty}^{\infty}d\eta_\alpha \, \text{cosh}\left(\eta_\alpha-y\right)
f(T_{FO}, (p\cdot u)_\alpha,\Pi_\alpha,\pi^{\mu\nu}_\alpha)   \right. \nonumber \\
& & + \left.  \left[p^x n_x^{(\alpha)} +p^y n_y^{(\alpha)}\right]
\int_{-\infty}^{\infty}d\eta_\alpha 
\,f(T_{FO}, (p\cdot u)_\alpha,\Pi_\alpha,\pi^{\mu\nu}_\alpha)\right\}
\end{eqnarray}
where we can then substitute in
\begin{equation}
\left(q_{\nu}\right)_{\alpha}=\frac{\left(n_{\nu}\right)_{\alpha}}{\left(n\cdot u\right)_{\alpha}}\frac{\nu_{\alpha}}{\sigma_{\alpha}}
\end{equation}
such that
\begin{eqnarray}
\frac{dN}{dyd^2p_T}& =& \frac{g}{(2\pi)^3}\,\sum_{\alpha=1}^{N_{SPH}}\left\{  m_{\perp} q_{0\,\alpha} 
\int_{-\infty}^{\infty}d\eta_\alpha \, \text{cosh}\left(\eta_\alpha-y\right)
f(T_{FO}, (p\cdot u)_\alpha,\Pi_\alpha,\pi^{\mu\nu}_\alpha)   \right. \\
& & + \left.  \left(\mathbf{p}_T\cdot \mathbf{q}_T\right)_\alpha
\int_{-\infty}^{\infty}d\eta_\alpha 
\,f(T_{FO}, (p\cdot u)_\alpha,\Pi_\alpha,\pi^{\mu\nu}_\alpha)\right\}\nonumber\\
& =& \frac{g}{(2\pi)^3}\,\sum_{\alpha=1}^{N_{SPH}}\,%
\left[ q_{0 \,\alpha} \,\mathcal{I}_1(\alpha,m,T_{FO})-(\mathbf{p}_T \cdot 
\mathbf{q}_{T})_\alpha\, \mathcal{I}_2(\alpha,m,T_{FO}) \right]\,
\end{eqnarray}
where

\begin{eqnarray}
\mathcal{I}_1(\alpha,m,T_{FO})&=&  m_{\perp} \int_{-\infty}^{\infty}d\eta_\alpha \, \text{cosh}\left(\eta_\alpha-y\right)
f(T_{FO}, (p\cdot u)_\alpha,\Pi_\alpha,\pi^{\mu\nu}_\alpha)    \nonumber\\
\mathcal{I}_2(\alpha,m,T_{FO})&=&   \int_{-\infty}^{\infty}d\eta_\alpha 
\,f(T_{FO}, (p\cdot u)_\alpha,\Pi_\alpha,\pi^{\mu\nu}_\alpha) \nonumber\\
\end{eqnarray}

We can now insert the distribution function in Eq.\ (\ref{eqn:cordiswithns})
\begin{eqnarray}
\mathcal{I}_1(\alpha,m,T_{0})&=& 
 m_{\perp}\sum_{n=0}^{\infty}\left(-a_\alpha\right)^n \lambda^{n+1}_\alpha  \int_{-\infty}^{\infty}d\eta_\alpha \, \text{cosh}\left(\eta_\alpha-y\right) e^{-(n+1)m_{\perp} \frac{u_{\tau}^{(\alpha)}}{T_0} \text{cosh}\left(\eta_\alpha-y\right)} \nonumber \\
&\times & \left[1+  \left( n+1\right) \left(\delta f^{(\alpha)Bulk}_{p}+\delta f^{(\alpha)Shear}_{p}\right)\right]   \nonumber\\
\mathcal{I}_2(\alpha,m,T_{0})&=&   
\,\sum_{n=0}^{\infty}\left(-a_\alpha\right)^n \lambda^{n+1}_\alpha \int_{-\infty}^{\infty}d\eta_\alpha\,  e^{-(n+1)m_{\perp} \frac{u_{\tau}^{(\alpha)}}{T_0} \text{cosh}\left(\eta_\alpha-y\right)}\nonumber \\
&\times &\,\left[1+  \left( n+1\right) \left(\delta f^{(\alpha)Bulk}_{p}+\delta f^{(\alpha)Shear}_{p}\right)\right] \nonumber\\
\end{eqnarray}
but we already know a portion of this from the combination of the ideal and bulk in \cite{Noronha-Hostler:2013gga}, which we will refer to here as $I_1^{\alpha+b}$ and $I_2^{\alpha+b}$, so
\begin{eqnarray}
\mathcal{I}_1(\alpha,m,T_{FO})&=& I_1^{\alpha+b}+ m_{\perp}
\sum_{n=0}^{\infty}\left( n+1\right) \left(-a_\alpha\right)^n \lambda^{n+1}_\alpha  \int_{-\infty}^{\infty}d\eta_\alpha \, \text{cosh}\left(\eta_\alpha-y\right) e^{-(n+1)m_{\perp} u_{\tau}^{(\alpha)}/T_0 \text{cosh}\left(\eta_\alpha-y\right)} \nonumber \\ &\times& \delta f^{(\alpha)Shear}_{p}\ \nonumber\\
\mathcal{I}_2(\alpha,m,T_{FO})&=&   I_2^{\alpha+b}+
\,\sum_{n=0}^{\infty}\left( n+1\right) \left(-a_\alpha\right)^n \lambda_\alpha^{n+1} \int_{-\infty}^{\infty}d\eta_\alpha\,  e^{-(n+1)m_{\perp} u_{\tau}^{(\alpha)}/T_0 \text{cosh}\left(\eta_\alpha-y\right)}\delta f^{(\alpha)Shear}_{p}\  \nonumber\\
\end{eqnarray}

\subsection{Details about $\delta f$ for shear}


The correction term from shear viscosity effects for a given particle species is
\begin{eqnarray}
\delta f^{(i)Shear}_{p}&=&\frac{1}{2s_0 \,T_0^3}\pi^{\mu\nu}p_{\mu}p_{\nu}\,
\end{eqnarray}
where $s_0$ is the entropy density at freeze-out.

Using the properties of the shear stress tensor we find, in explicit form,
\begin{eqnarray}
\pi^{\mu\nu}p_{\mu}p_{\nu}&=& m_{\perp}^2 \left[\pi^{00} \text{cosh}^2(\eta- y) +\tau^2\pi^{33}  \text{sinh}^2(\eta- y) \right]+p^2_x\pi^{11}+p_y^2\pi^{22}+2p_x p_y \pi^{12}
\end{eqnarray}

Substituting that expression in the shear correction term, one obtains for each SPH particle
\begin{eqnarray}
\mathcal{I}_1(\alpha,m,T_{0})&=& I_1^{\alpha+b}+
\frac{m_{\perp}}{2s_0 T_0^3}\sum_{n=0}^{\infty}\left( n+1\right) \left(-a_\alpha\right)^n \lambda_\alpha^{n+1}  \int_{-\infty}^{\infty}d\eta_\alpha \, \text{cosh}\left(\eta_\alpha-y\right) \nonumber \\ &\times& e^{-(n+1)m_{\perp} u_{\tau}^{(\alpha)}/T_0 \text{cosh}\left(\eta_\alpha-y\right)} \pi_\alpha^{\mu\nu}p_{\mu}p_{\nu}\nonumber\\
\mathcal{I}_2(\alpha,m,T_{0})&=&   I_2^{\alpha+b}+
\frac{1}{2s_0 T_0^3}\sum_{n=0}^{\infty}\left( n+1\right) \left(-a_\alpha\right)^n \lambda_\alpha^{n+1} \int_{-\infty}^{\infty}d\eta_\alpha  e^{-(n+1)m_{\perp} u_{\tau}^{(\alpha)}/T_0 \text{cosh}\left(\eta_\alpha-y\right)}\pi_\alpha^{\mu\nu}p_{\mu}p_{\nu}\,. \nonumber \\
\end{eqnarray}

After some manipulations, our final equations become
\begin{eqnarray}
\mathcal{I}_1(\alpha,m,T_{0})&=& I_1^{\alpha+b}+
\frac{1}{s_0 T_0^3}\sum_{n=0}^{\infty}\left( n+1\right) \left(-a_\alpha\right)^n \lambda_\alpha^{n+1} \nonumber\\
& \times & \left\{ \left(E\left[p^2_x\pi_\alpha^{11}+p_y^2\pi_\alpha^{22}+2p_x p_y \pi_\alpha^{12}\right]+\frac{1}{4} E^3 \left[3\pi_\alpha^{00}-\tau^2\pi_\alpha^{33}\right] \right)K_1\left((n+1)\frac{E u_{\tau}^{(\alpha)}}{T_0}\right)\right.+\nonumber\\
& & +\left.\frac{1}{4} m_{\perp}^3\left[ \pi_\alpha^{00}+\tau^2\pi_\alpha^{33} \right] K_3\left((n+1)\frac{m_{\perp} u_{\tau}^{(\alpha)}}{T_0}\right) \right\} \nonumber\\
\mathcal{I}_2(\alpha,m,T_{0})&=& I_2^{\alpha+b}+
\frac{1}{s_0 T_0^3}\sum_{n=0}^{\infty}\left( n+1\right) \left(-a_\alpha\right)^n \lambda_\alpha^{n+1} \nonumber\\
&\times & \left\{\left(    p^2_x\pi_\alpha^{11}+p_y^2\pi_\alpha^{22}+2p_x p_y \pi_\alpha^{12}+\frac{1}{2}m_{\perp}^2\left[\pi_\alpha^{00}-\tau^2\pi_\alpha^{33}\right]\right)
K_0\left((n+1) \frac{m_{\perp} \gamma_\alpha}{T_0}\right)\right.+\nonumber\\
&\times & +\left.\frac{1}{2}m_{\perp}^2 \left(\pi_\alpha^{00}+\tau^2\pi_\alpha^{33}\right)K_2\left((n+1) \frac{ m_{\perp} \gamma_\alpha}{T_0}\right)\right\} \nonumber\\
\end{eqnarray}
where $K_\beta(x)$ is the modified Bessel function.

\section{Normal vector of isothermal surface and the SPH formalism}\label{normalvector}

The normalized normal vector to the isothermal surface is
\begin{equation}
n_\mu = \frac{\left(\partial_\tau T,\partial_x T, \partial_y T\right)}{\sqrt{\left(\partial_\tau T\right)^2-\left(\partial_x T\right)^2-\left(\partial_y T\right)^2}}\,.
\end{equation}
Since in the SPH method the spatial gradients of the pressure are known \cite{Hama:2004rr} and, using the Gibbs-Duhem relation $\partial_\mu T = \partial_\mu P/s$, we just need to determine $\partial_\tau T$ to obtain $n_\mu$. Using that $DP = \gamma \partial_\tau P + \left(\mathbf{u}\cdot \nabla P\right)$, $dP/d\varepsilon=c_s^2$ hence $DP = c_s^2 \,D\varepsilon$, and the energy conservation equation
\begin{equation}
D\varepsilon + (\varepsilon+P+\Pi)\theta - \pi_{\mu\nu}\sigma^{\mu\nu}=0
\end{equation}
we find that
\begin{equation}
\partial_\tau P = \frac{1}{\gamma} \left[-c_s^2\,\theta \left(\varepsilon+P+\Pi   \right)+c_s^2\, \pi_{\mu\nu}\sigma^{\mu\nu}- \left(\mathbf{u}\cdot \nabla P\right)\right]
\end{equation}
and thus
\begin{equation}\label{eqn:normtau}
\partial_\tau T = \frac{1}{\gamma\,s} \left[-c_s^2\,\theta \left(\varepsilon+P+\Pi   \right)+c_s^2\, \pi_{\mu\nu}\sigma^{\mu\nu}- \left(\mathbf{u}\cdot \nabla P\right)\right]\,.
\end{equation}

Both $\eta/s$ and $\zeta/s$ are still relatively small in all of our calculations, which means that the contributions from the bulk pressure and shear stress tensor in Eq.\ (\ref{eqn:normtau}) are still in general very small compared to that from the energy density and pressure ($\varepsilon+P\approx 1.5$ whereas the components of $\pi^{\mu\nu}\approx10^{-3}$ and $\Pi\approx10^{-3}-10^{-1}$). This then means that the primary contribution to the uncorrected flow harmonics comes from the viscous corrected flow and not from the details of the freeze-out hypersurface (which remains very similar to the one found in the ideal hydro case).

\input{bibi1}

\end{document}